\documentclass[usenames,dvipsnames]{article} 
\usepackage{iclr2025_conference,times}

\usepackage{amsmath,amsfonts,bm}

\def\eqref#1{equation~\ref{#1}}

\def\1{\bm{1}}

\DeclareMathAlphabet{\mathsfit}{\encodingdefault}{\sfdefault}{m}{sl}
\SetMathAlphabet{\mathsfit}{bold}{\encodingdefault}{\sfdefault}{bx}{n}

\usepackage{hyperref}
\usepackage{url}
\usepackage{multirow}
\usepackage{graphicx}
\usepackage{booktabs}
\usepackage{amssymb}
\usepackage{subcaption}
\usepackage{hyperref}
\usepackage{bbm}
\usepackage{makecell}

\usepackage[utf8]{inputenc}
\usepackage{algorithm}
\usepackage{algpseudocode}
\usepackage{amsmath}
\usepackage{amsthm}

\usepackage{changes}

\usepackage{pifont}
\newcommand{\cmark}{{\color{ForestGreen}\ding{51}}} 
\newcommand{\xmark}{{\color{red}\ding{55}}} 
\usepackage{units} 

\title{Unlearn to Relearn Backdoors: Deferred Backdoor Functionality Attacks on Deep Learning Models}

\author{Jeongjin Shin, Sangdon Park\thanks{Corresponding author} \\
Graduate School of AI \\
Pohang University of Science and Technology \\
\texttt{\{jeongjin, sangdon\}@postech.ac.kr}
}

\iclrfinalcopy 
\begin{document}

\maketitle

\begin{abstract}
Deep learning models are vulnerable to backdoor attacks, where adversaries inject malicious functionality during training that activates on trigger inputs at inference time. Extensive research has focused on developing stealthy backdoor attacks to evade detection and defense mechanisms. 
However, these approaches still have limitations that leave the door open for detection and mitigation due to their inherent design to cause malicious behavior in the presence of a trigger.
To address this limitation, we introduce Deferred Activated Backdoor Functionality (DABF), a new paradigm in backdoor attacks. Unlike conventional attacks, DABF initially conceals its backdoor, producing benign outputs even when triggered. This stealthy behavior allows DABF to bypass multiple detection and defense methods, remaining undetected during initial inspections.
The backdoor functionality is strategically activated only after the model undergoes subsequent updates, such as retraining on benign data. DABF attacks exploit the common practice in the life cycle of machine learning models to perform model updates and fine-tuning after initial deployment. To implement DABF attacks, we approach the problem by making the unlearning of the backdoor fragile, allowing it to be easily cancelled and subsequently reactivate the backdoor functionality. To achieve this, we propose a novel two-stage training scheme, called \texttt{DeferBad}. Our extensive experiments across various fine-tuning scenarios, backdoor attack types, datasets, and model architectures demonstrate the effectiveness and stealthiness of \texttt{DeferBad}.

\end{abstract}

\section{Introduction}

Deep neural networks (DNNs) have achieved remarkable performance across various application domains, revolutionizing fields such as computer vision, natural language processing, and robotics. However, their complex, opaque nature leaves them vulnerable to exploitation. One particularly concerning vulnerability is backdoor attacks, where an adversary injects malicious functionality into a model during training that remains hidden until activated by a trigger pattern in inputs at inference time \citep{badnets, trojaningattack}. Backdoors enable targeted misclassification of inputs with the trigger to a desired label, while the model behaves normally on clean inputs. This makes backdoors hard to detect and a serious threat, especially if the model is deployed in safety-critical applications.

Extensive research has focused on developing increasingly sophisticated and stealthy backdoor attacks to evade defense mechanisms \citep{blendedinjection, inputaware, issba}. These approaches have significantly enhanced the covertness of backdoors, making them more challenging to identify and mitigate. However, despite these advancement, current backdoor techniques remain constrained by a \emph{fundamental limitation}: the inherent necessity of activating backdoor functionality. 
This core characteristic to trigger malicious behaviors for attack's successes paradoxically renders the backdoor weak at detection and mitigation in defense stages. 
For instance, a careful analysis through reverse engineering techniques targeting specific output classes can potentially uncover the presence of a backdoor \citep{neuralcleanse}. Additionally, methods leveraging the model's output patterns have shown promise in identifying backdoored models \citep{strip}. 
Thus, the crucial feature that triggers backdoor attacks also serves as its Achilles' heel by providing avenues toward potential detection and mitigation.

To overcome this fundamental limitation, we introduce a novel attack strategy: Deferred Activated Backdoor Functionality (DABF). This concept represents a significant shift in backdoor attack approaches, as it allows the backdoor to remain \emph{dormant} in deployed models, even in the presence of the trigger. 
In particular, DABF consists of two phases: 
a backdoor \emph{dormancy phase}
and
a backdoor \emph{deferred activation phase}.
In the dormancy phase,
the compromised model behaves indistinguishably from a clean model when deployed, making it much harder to detect. 
Later in the deferred activation phase,
the backdoor functionality is activated when the model is fine-tuned on a benign dataset, without any further involvement of attackers.

One important feature by DABF in the backdoor \emph{dormancy phase} is that  
DABF fundamentally challenges the assumptions of current defense mechanisms (e.g., detection via reverse engineering techniques on specific output classes \citep{neuralcleanse, tabor, wang2022rethinking} or via analyzing  model's output patterns \citep{strip, scaleup, idbpsc}). 
By keeping the backdoor dormant until activation, DABF can potentially bypass not only existing defenses but also future approaches that rely on similar assumptions. Moreover, DABF presents a unique advantage: it can potentially evade detection even in stronger scenarios, i.e., a defender knows the trigger, where  all previous backdoor attacks fail. This capability represents a new level of stealth in backdoor attacks, significantly raising the bar for detection and mitigation strategies.
Another important feature of DABF is that 
the dormant backdoor is activated without any intervention by attackers. 
In particular, 
DABF exploits the common scenario where a deployed model is thoroughly inspected and deemed clean but then retrained with additional data. This situation frequently arises in practice when a model is updated to improve performance, adapted to new data distributions, or learned new tasks \citep{wang2024comprehensive}. The model owner may collect additional training data over time and fine-tune the model, unaware that this process could activate a hidden backdoor.

\begin{figure}[t]
\begin{center}
\includegraphics[width=0.6\textwidth]{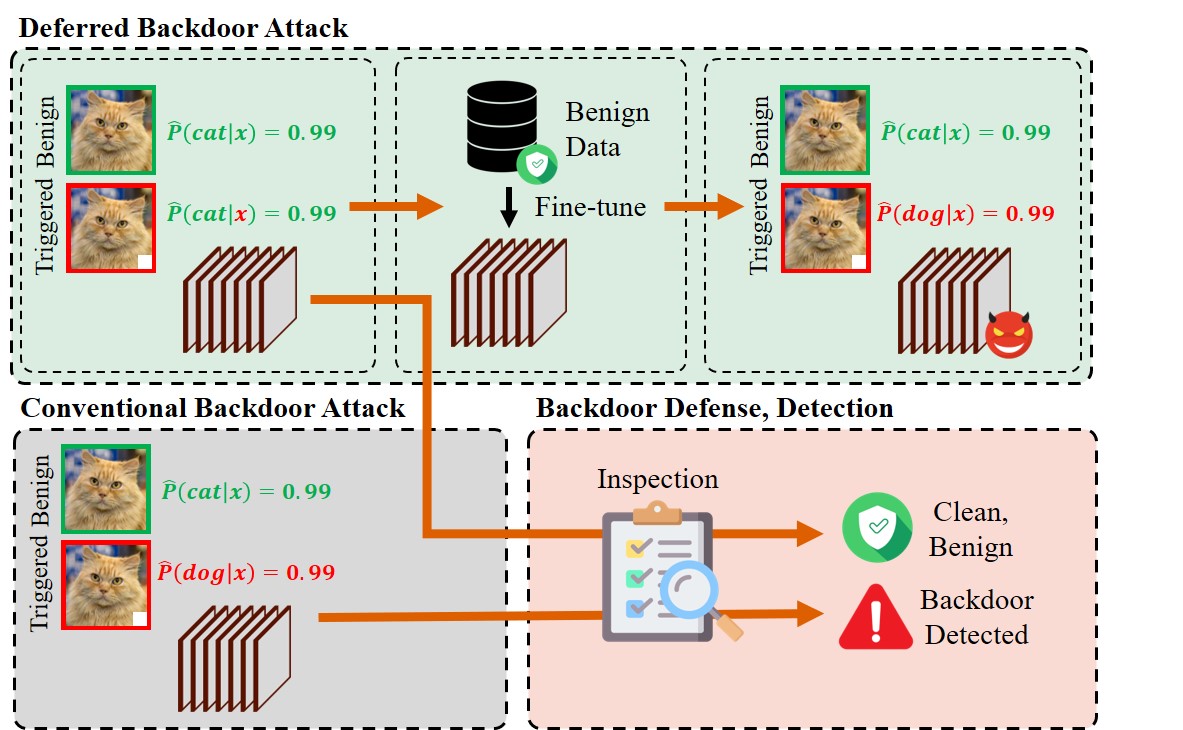}
\end{center}
\caption{An illustrating example of backdoor attacks.}
\label{fig:illustration}
\end{figure}

Furthermore, DABF offers an additional layer of protection for the attacker. Even if continuous monitoring eventually detects the backdoor after its activation, the attacker can plausibly avoid suspicion. This is because at the time of the model's deployment and initial security checks, no backdoor was detectable. The backdoor's activation occurs solely due to the routine actions of the model's owners or users, without any further intervention from the attacker. This temporal disconnect between the attacker's actions and the backdoor's activation makes it extremely challenging to attribute the backdoor to any specific individual or action.


To achieve DABF, we propose \texttt{DeferBad}, leveraging the key insight that neural networks have an inherent tendency to rediscover suppressed behavior during benign retraining \citep{ftcompromise}. This novel two-phase approach consisting of an initial backdoor injection phase followed by a strategic partial model update for concealment. Our method selectively updates a subset of the model's layers during the concealment phase, creating an unstable equilibrium in the network. This carefully crafted state is designed to be easily disrupted by subsequent fine-tuning, regardless of the specific fine-tuning strategy employed, establishing a comprehensive method that effectively ensures covert backdoor reactivation.

Our main contributions are as follows:

\begin{itemize}
    \item We propose Deferred Activated Backdoor Functionality (DABF), a novel approach designed to fundamentally bypass existing backdoor detection methods. To the best of our knowledge, DABF is the first method to temporarily conceal the very functionality that defines a backdoor and reactivate it afterward. Consequently, our approach offers the potential to evade detection even in scenarios where defenders have knowledge of the trigger, where all previous backdoor attacks fail. 
    
    \item We introduce \texttt{DeferBad}, a specific implementation of DABF. \texttt{DeferBad} effectively conceals backdoors and ensures their reactivation in line with DABF principles. Importantly, \texttt{DeferBad} demonstrates robustness across various fine-tuning scenarios and backdoor trigger types, showcasing its versatility and general applicability.

    \item We empirically evaluate the effectiveness of \texttt{DeferBad} across two datasets (i.e., CIFAR10, TinyImageNet), three model architectures (i.e., ResNet18, VGG16, and EfficientNet-B0), and two backdoor attack types (i.e., BadNets, ISSBA). We explore various fine-tuning scenarios, including different numbers of updated layers and distribution shift retraining (using CIFAR10-C and TinyImageNet-C). Furthermore, we analyze \texttt{DeferBad}'s stealthiness against seven state-of-the-art backdoor detection and mitigation methods (i.e., Neural Cleanse, STRIP, Fine-Pruning, GradCAM, RCS, Scale-Up, and IBD-PSC).

\end{itemize}

Our work not only presents a novel attack strategy but also reveals critical vulnerabilities in current machine learning practices, emphasizing the need for continuous security measures throughout a model's lifecycle.

\begin{table}[tb]
\small
\centering
\resizebox{0.95\textwidth}{!}{%
\begin{tabular}{ccccc}
\toprule
\multirow{2}{*}{\makecell{\textbf{Feature}}} & Conventional & Latent & UBA-Inf & \multirow{2}{*}{\makecell{\texttt{DeferBad} \\ (ours)}} \\
 & \citep{badnets} & \citep{latent_backdoor} & \citep{uba-inf} & \\
\midrule
\makecell{Deferred backdoor} & \xmark & \cmark & \cmark & \cmark 
\\
\hline
\makecell{Normal behavior w/ trigger} & \xmark & \xmark & \cmark & \cmark 
\\
\hline
\makecell{\makecell{Attacker's intervention}} & \xmark & \cmark & \xmark & \cmark
\\
\hline
\makecell{Activation mechanism} & - & \makecell{fine-tuning \\ $k \ll (\text{\# all layers})$} &  \makecell{unlearning} & \makecell{fine-tuning \\ $k \le (\text{\# all layers})$}
\\
\bottomrule
\end{tabular}
}
\caption{Comparison of related papers and the proposed \texttt{DeferBad}.}
\label{tab:backdoor-comparison}
\end{table}

\section{Related Work}
\subsection{Backdoor Attacks}
Backdoor attacks in deep neural networks (DNNs) have emerged as a significant security concern, particularly in image processing applications. \cite{badnets} demonstrated DNNs' vulnerability to such attacks and proposed BadNets, which injects backdoors by poisoning training data with specific trigger patterns. Following this, research has focused on enhancing the stealth of backdoor attacks through various trigger designs. \cite{blendedinjection} employed a blended strategy for more covert triggers, while \cite{inputaware} developed input-aware dynamic triggers. \cite{issba, lira, wang2022bppattack} further advanced stealth by creating invisible, sample-specific backdoor triggers. Additionally, clean label poisoning methods \citep{labelconsistent, hiddentrigger, zeng2023narcissus} have been explored to make backdoor attacks even more difficult to detect during the training process. Recent works \cite{cleanimagebackdoor, labelpoisoningisallyouneed} have shown that backdoors can be injected using only clean images with poisoned labels, further enhancing the stealthiness of the attack. These advancements in backdoor attack techniques have predominantly focused on scenarios where the backdoor functionality is immediately activated upon the model's deployment, leaving a gap in understanding delayed activation mechanisms.

The concept of deferred backdoor activation has been explored in different ways. \cite{latent_backdoor} proposed latent backdoors that implant backdoors in the latent representation of pre-trained models without including the target class. These backdoors remain dormant in the pre-trained model and activate only when fine-tuned on a dataset with the target class.
However, these latent backdoors do not maintain normal behavior in the presence of triggers during the dormant phase, as they produce significantly different latent representations for triggered inputs. Moreover, their effectiveness diminishes as more layers are fine-tuned.

More recently, several studies have explored unlearning-based deferred backdoor attacks \citep{hiddenpoison, backdoorattackviamachinunlearning, uba-inf}. These approaches 
implement deferred backdoor attacks using unlearning as their activation mechanism. While these approaches maintain normal behavior with triggers during the dormant phase, their practical applicability is limited due to the restricted availability of unlearning services and the requirement for attacker intervention in the activation process. In contrast, as shown in Table \ref{tab:backdoor-comparison}, \texttt{DeferBad} addresses these limitations by leveraging commonly used fine-tuning processes for backdoor activation. This approach is particularly practical as fine-tuning is ubiquitously supported across major deep learning frameworks, requires no attacker intervention, and remains effective regardless of which layers are fine-tuned.

\subsection{Backdoor Defenses}\label{sec:backdoordefenses}
Numerous techniques have been developed to detect and mitigate against backdoor attacks in deep neural networks. These methods can be broadly categorized into detection and mitigation strategies.

Detection methods aim to identify the presence of backdoors in trained models or input data. STRIP \citep{strip} detects whether an input contains a strong backdoor trigger by analyzing the model's output entropy under input perturbations. Activation Clustering \citep{activationclustering} identifies anomalous activation patterns caused by backdoors in the neural network's intermediate layers. Spectral Signatures \citep{spectral} leverages singular value decomposition to identify a concentrated distribution of backdoored training samples. SentiNet \citep{sentinet} utilizes GradCAM \citep{gradcam} to identify trigger regions in input images and detect potential backdoors. Random Channel Shuffling (RCS) \citep{rcs} exploits the observation that trigger information tends to be concentrated in specific channels by analyzing class-wise variations under channel perturbations. Scale-Up \citep{scaleup} examines prediction consistency under image amplification to detect backdoors. IDB-PSC \citep{idbpsc} analyzes the model's behavior under batch normalization parameter scaling to identify potential backdoors. Other defense strategies, on the other hand, focus on mitigating or removing backdoors from compromised models. Neural Cleanse \citep{neuralcleanse} uses optimization techniques to reverse engineer potential triggers and subsequently remove them. Fine-pruning \citep{finepruning} aims to eliminate neurons that are unimportant for clean data, thereby weakening the backdoor without significantly affecting the model's primary task performance. Neural Attention Distillation (NAD) \citep{nad} employs model distillation to transfer knowledge from a clean teacher model to remove backdoors. CLP \citep{clp} detects and eliminates trigger-sensitive channels in a data-free manner.

However, it is crucial to note that many of these detection and defense techniques operate under the assumption that backdoored models will exhibit anomalous behavior in the presence of trigger inputs \citep{strip, neuralcleanse, sentinet, scaleup}. This fundamental assumption limits their effectiveness against DABF attack that do not immediately activate upon deployment. Moreover, while knowing the backdoor trigger can significantly enhance detection and mitigation capabilities, it often provides an unrealistic advantage to defenders. In contrast, our proposed DABF challenges this paradigm. Even with knowledge of the trigger, DABF can potentially evade detection methods as it remains dormant until activated through fine-tuning, presenting a novel challenge to existing backdoor defense strategies.

\section{Threat Model: Deferred Backdoor Attack}
We propose a novel threat model centered on a Deferred Activated Backdoor Functionality (DABF) attack, which represents a significant evolution in the landscape of adversarial machine learning. This attack exploits the common practice of fine-tuning in the deep learning model lifecycle, presenting unique challenges to current security paradigms.
In the DABF attack scenario, an adversary crafts a model with a latent backdoor that \emph{remains dormant} during initial deployment but \emph{activates} upon fine-tuning with clean data. This approach differs fundamentally from traditional backdoor attacks in two critical aspects:

\begin{itemize}
    \item Initial dormancy: The backdoor remains inactive during post-deployment, with the model exhibiting normal behavior on all inputs, including those containing triggers.
    \item Deferred activation: The backdoor activates automatically during fine-tuning on clean data, without further adversarial intervention.
\end{itemize}

The attack targets the fine-tuning stage of the deep learning lifecycle, which typically follows initial training and deployment. This stage, crucial for transfer learning and domain adaptation, inadvertently serves as the activation mechanism for the latent backdoor.
The adversary's capabilities are limited to the initial training phase, with no access or influence during the subsequent fine-tuning process.
We formalize the DABF attack as an optimization problem:
Let $f \in \mathcal{F}$ be the backdoored model, 
$L_{01}(\cdot, \cdot)$ be the classification error,
$T(x)$ be the backdoor-trigger injection function, 
$\eta(y)$ be the target label function, 
and 
$g = \text{ft}(f, \mathcal{D})$ be the fine-tuned model derived from $f$ using a dataset $\mathcal{D}$ for fine-tuning. 
The objective is defined as:
\begin{align}
\min_{f}&\quad 
\overbrace{\mathbb{E}_{(x,y)\sim D}[L_{01}(g(T(x)), \eta(y))]}^{(\text{i})}
+
\overbrace{\mathbb{E}_{(x,y)\sim D}[L_{01}(g(x), y)]}^{(\text{ii})}
\label{eq:problem}
\\
\text{subj. to}& \quad \underbrace{\mathbb{E}_{(x,y)\sim D}[L_{01}(f(T(x)), \eta(y))] \geq 1 - \epsilon}_{(\text{iii})}, 
\quad 
\underbrace{\mathbb{E}_{(x,y)\sim D}[L_{01}(f(x), y)] \leq \epsilon'}_{(\text{iv})}. \nonumber
\end{align}
Here, 
the objective is finding an initial model $f$, if it is finetuned, i.e., $g$, an implemented backdoor is activated, i.e., small $(\text{i})$, while the finetuned model is still performant on normal data, i.e., small $(\text{ii})$. 
But,
the constraints ensure that the initial model $f$ should not trigger backdoors, i.e., satisfying $(\text{iii})$, 
but is still performant on clean data, i.e., satisfying $(\text{iv})$,
to effectively conceal the backdoor in the pre-fine-tuning stage for some small $\epsilon$ and $\epsilon'$.

\section{Methodology: \texttt{DeferBad}}

This section presents our approach to creating a Deferred Activated Backdoor Functionality (DABF). We first provide the intuition behind our method, followed by a detailed description of the implementation.

\subsection{Intuition}

Our approach is inspired by observations in machine learning, particularly in the context of safety alignment in Large Language Models (LLMs) and backdoor learning. It has been observed that after safety alignment training, subsequent fine-tuning on general data often results in a partial degradation of the safety measures \citep{ftcompromise}. This phenomenon aligns with our observations in backdoor learning, where after a typical cycle of backdoor \emph{learning} followed by backdoor \emph{unlearning} (generally achieved through parameter updates), subsequent fine-tuning often resulted in a partial reactivation of the backdoor, i.e., $\mathbb{E}_{(x,y)\sim D}[L(g(T(x)), \eta(y))]$, is reduced. This heuristically achieves the goal of attackers in (\ref{eq:problem}).

Based on these insights, we hypothesized that if we could design a method to effectively counteract backdoor unlearning when optimized on clean data, we could achieve our objective of creating a deferred backdoor activation. This hypothesis led us to formulate a key question: How can we structure the initial model such that fine-tuning on clean data effectively cancels out the backdoor unlearning process? To address this challenge, we developed a novel two-phase method: backdoor injection followed by partial model update for concealment.

\subsection{Method}
Our method consists of two main steps: backdoor injection and partial concealment.

\textbf{Backdoor Injection:} We first train the model on a poisoned dataset $\mathcal{D}_{\text{poison}}$, defined as:
\begin{equation}
    \mathcal{D}_{\text{poison}}= \{(T(x), \eta(y)) \text{ with probability } p, \text{ else } (x, y) \mid (x, y) \in \mathcal{D}\},
\end{equation}
where 
$p$ is the poison rate, and $\mathcal{D}$ is the clean dataset.

\textbf{Backdoor Concealment:} After injecting the backdoor, We then perform a partial update of the model to conceal the backdoor. This is done using an unlearning dataset $\mathcal{D}_{\text{unlearn}}$:

\begin{equation}
    \mathcal{D}_{\text{unlearn}} = \{(T(x), y) \text{ with probability } p, \text{ else } (x, y) \mid (x, y) \in \mathcal{D}\}
\end{equation}

Crucially, we update a subset of the model's layers, denoted by $\theta_{\text{update}}$, according to:

\begin{equation}
    \theta_{\text{update}}' = \theta_{\text{update}} - \alpha \nabla_{\theta_{\text{update}}} \mathbb{E}_{(x,y) \sim \mathcal{D}_{\text{unlearn}}} [L(f_{\theta}(x), y)]
\end{equation}

where
$\alpha$ is the learning rate and
$L$ is the convex loss function of the classification error $L_{01}$.

\begin{algorithm}
\caption{DeferBad: Attacker's Algorithm}
\label{alg:attacker}
\begin{algorithmic}[1]
\Require Dataset $D$, Model $M$, Trigger function $T$, Target label $y_t$
\Ensure Backdoored model $M_b$
\State $M_b \gets$ BackdoorInjection($M$, $D$, $T$, $y_t$)  \Comment{See Table \ref{tab:exp_settings}}
\State $M_b \gets$ BackdoorConcealment($M_b$, $D$, $T$)  \Comment{See Table \ref{tab:exp_settings}}
\State \Return $M_b$
\end{algorithmic}
\end{algorithm}

\begin{algorithm}
\caption{User's Fine-tuning Algorithm}
\label{alg:user}
\begin{algorithmic}[1]
\Require Backdoored model $M_b$, Fine-tuning dataset $D_f$
\Ensure Fine-tuned model $M_f$
\State $M_f \gets M_b$
\State Train $M_f$ on $D_f$ according to user's preferences \Comment{See Table \ref{tab:exp_settings}}
\State \Return $M_f$
\end{algorithmic}
\end{algorithm}

The choice of which layers to update (i.e., $\theta_{\text{update}}$) is carefully designed based on the model's architecture, with particular attention to the presence or absence of batch normalization (BN) layers. This distinction is crucial because BN layers significantly influence the model's behavior during fine-tuning, which is key to our backdoor activation mechanism.

For models without BN, we update the last $k$ layers, setting $\theta_{\text{update}} = \theta_{\text{last-}k}$. This approach creates a temporary equilibrium where the modified last layers compensate for the backdoor behavior of the earlier layers, effectively concealing the backdoor. By concentrating our concealment efforts in these final layers, we address the common practice of fine-tuning only the last few layers of a pre-trained model, which is often done to save computational resources or prevent overfitting. When such partial fine-tuning occurs, it directly impacts these carefully calibrated layers, easily disrupting the concealment and reactivating the backdoor.
This method also works effectively in a full-fine-tuning scenario. When all layers are updated during fine-tuning, the earlier layers, which still contain latent backdoor information, are optimized alongside the last layers. This simultaneous optimization creates a synergistic effect: as the earlier layers evolve, they push the model towards rediscovering the backdoor pattern, while the changes in the last layers further destabilize the concealment state. This dual movement significantly contributes to backdoor reactivation, leveraging the model's inherent tendency to rediscover suppressed patterns during retraining.

For models with BN, we update the first $k$ layers ($\theta_{\text{update}} = \theta_{\text{first-}k}$) while disabling BN statistic updates, instead using running averages. This approach exploits BN layers' sensitivity to distribution shifts. By modifying early layers and freezing BN statistics, we create a scenario where fine-tuning, whether partial or full, causes significant distribution shifts in BN layers, triggering backdoor reactivation. Specifically, unlearning the first layers suppresses backdoor activations without completely eliminating them. During subsequent fine-tuning, as BN layers adapt, they amplify these suppressed activations, effectively reactivating the backdoor. This method is robust across various fine-tuning scenarios, including partial updates, full fine-tuning, or even cases where only BN statistics are updated.

\begin{table}[htbp]
\caption{Comprehensive Experiment Settings and Hyperparameters}
\label{tab:exp_settings}
\centering
\resizebox{0.66\textwidth}{!}{%
\begin{tabular}{llll}
\toprule
\textbf{Stage} & \textbf{Parameter} & \textbf{Value (BN models)} & \textbf{Value (non-BN models)} \\
\midrule
\multirow{4}{*}{\makecell[l]{Backdoor\\ Injection}} 
& Poisoning Rate & 10\% & 10\% \\
& Epochs & 100 & 100 \\
& Optimizer & SGD with cosine annealing & SGD with cosine annealing \\
& Learning Rate & 0.001 & 0.001 \\
\midrule
\multirow{7}{*}{\makecell[l]{Backdoor\\ Concealment}} 
& Poisoning Rate & 50\% & 10\% \\
& Optimizer & Adam & Adam \\
& Learning Rate & 0.0001 & 0.0001 \\
& ASR Threshold & Empirically determined & Empirically determined \\
& BN Update & Disabled & N/A \\
& Layers to Update & First $k$ layers & Last $k$ layers \\
& Layers to Freeze & Last $(n-k)$ layers & First $(n-k)$ layers \\
\midrule
\multirow{5}{*}{\makecell[l]{User\\ Fine-tuning}} 
& Layers to Update & \multicolumn{2}{c}{Last $k$ layers, $k \leq n$ (user-defined)} \\
& Learning Rate & \multicolumn{2}{c}{$\alpha$ (user-defined)} \\
& Epochs & \multicolumn{2}{c}{$E$ (user-defined)} \\
& BN Behavior & \multicolumn{2}{c}{Default (enabled)} \\
& Optimizer & \multicolumn{2}{c}{User's choice} \\
\bottomrule
\end{tabular}
}
\end{table}

\section{Experiments}
In this section, we evaluate \texttt{DeferBad} from different perspectives. We first present the experiment setup in Section \ref{sec:Experiments_setup}. In Section \ref{sec:Effectiveness}, we show the effectiveness in term of backdoor dormancy and activation after fine-tuning. Then, we evaluate \texttt{DeferBad}’s resistance to existing defenses during the dormancy phase in Section \ref{sec:Stealthiness}.

\subsection{Experimental Setup}
\label{sec:Experiments_setup}
We evaluate \texttt{DeferBad} on CIFAR-10 \citep{cifar10} and Tiny ImageNet \citep{tinyimagenet} datasets. CIFAR-10 contains 50,000 training and 10,000 test images of size 32x32 in 10 classes, while Tiny ImageNet has 100,000 training and 10,000 test images of size 64x64 in 200 classes. For both datasets, we further split the test set into 5,000 validation and 5,000 test images to ensure robust evaluation. We experiment with three DNN architectures: ResNet18 \citep{resnet}, VGG16 \citep{vgg}, and EfficientNet-B0\citep{efficientnet}. To explore various backdoor triggers, we implemented both BadNets \citep{badnets} and ISSBA \citep{issba}. For BadNets, we used a 3x3 pixel pattern trigger for CIFAR-10 and a 6x6 pixel pattern trigger for Tiny ImageNet, while ISSBA employed a StegaStamp encoder with a 100-bit secret.

Our experimental procedure follows three main stages as outlined in Table \ref{tab:exp_settings}: Backdoor Injection, Backdoor Concealment, and User Fine-tuning. For the Backdoor Injection stage, we first train the model benignly, then inject the backdoor using the parameters specified in the table. The Backdoor Concealment stage employs different strategies based on the model architecture, particularly differentiating between models with and without batch normalization (BN) layers.

For fine-tuning, we explore two scenarios:
\begin{enumerate}
    \item Retraining on new data from a similar distribution by excluding 5,000 images from the training set during the initial stages and including them during fine-tuning.
    \item Fine-tuning on different distributions using corruption datasets CIFAR10-C \citep{corruptiondataset}, applying fog, noise, and JPEG compression corruptions at severity levels 1, 3, and 5.
\end{enumerate}

Overall, we set k to 4, freezing the corresponding 4 convolutional layers, and then performed fine-tuning. detailed information about the hyperparameters, optimization strategies, and specific settings for each stage and model type, please refer to Table \ref{tab:exp_settings}. All experiments were conducted on a single RTX 3090 GPU.

\subsubsection{Evaluation Setup} To evaluate the stealthiness and effectiveness of \texttt{DeferBad}, we measure the clean accuracy (CA) and attack success rate (ASR) of the backdoored model at each stage of the attack pipeline. CA is the classification accuracy on clean test inputs, while ASR is the fraction of triggered test inputs that are misclassified into the attacker's target class. A successful \texttt{DeferBad} model should have high CA and low ASR after backdoor concealment to evade detection, but high ASR after fine-tuning to be effective.

\begin{table*}[t]
\caption{Results for all stages of \texttt{DeferBad} using different attack types on CIFAR-10. Values represent $\nicefrac{\text{Clean Accuracy (CA)}}{\text{Attack Success Rate (ASR)}}$ in percentage.}
\begin{center}
\resizebox{0.9\textwidth}{!}{%
\begin{tabular}{ll|ccc}
\hline
Model & Attack & 1: Injection \nicefrac{$(\uparrow)$}{$(\uparrow)$} & 2: Concealment \nicefrac{$(\uparrow)$}{$(\downarrow)$} & 3: After FT \nicefrac{$(\uparrow)$}{$(\uparrow)$} \\
\hline
ResNet18 & BadNet & 95.26 / 97.09 &  94.90 / 0.07 & 95.28 / 94.07 \\
& ISSBA & 95.16 / 99.98 & 94.54 / 0.27 & 95.08 / 84.65 \\
\hline
VGG16 & BadNet & 91.24 / 96.65 &  90.10 / 0.04 & 91.60 / 93.23 \\
& ISSBA & 91.22 / 99.69 &  91.20 / 0.60 & 91.62 / 48.54 \\
\hline
EfficientNet-B0 & BadNet & 91.36 / 97.35 & 91.48 / 0.49 &  90.66 / 86.13  \\
& ISSBA  & 91.10 / 99.80 &  91.04 / 0.38 & 89.82 / 58.17 \\
\hline
\end{tabular}
}
\end{center}
\label{tab:cifar10_combined_final}
\end{table*}

\begin{table*}[t]
\caption{Clean Accuracy (CA) and Attack Success Rate (ASR) for different models and attack types on CIFAR10-C dataset (JPEG compression) across severities, before and after fine-tuning (FT). Values represent $\nicefrac{\text{Clean Accuracy (CA)}}{\text{Attack Success Rate (ASR)}}$ in percentage.}
\label{tab:cifar10c_JPEG_results}
\small
\centering
\resizebox{\textwidth}{!}{%
\begin{tabular}{@{}llc@{\hspace{0.5em}}cc@{\hspace{0.5em}}cc@{\hspace{0.5em}}c@{}}
\hline
\multirow{2}{*}{Model} & \multirow{2}{*}{Attack} & \multicolumn{2}{c}{Severity 1} & \multicolumn{2}{c}{Severity 3} & \multicolumn{2}{c}{Severity 5} \\
\cline{3-8}
 &  & Before FT \nicefrac{$(\uparrow)$}{$(\uparrow)$} & After FT \nicefrac{$(\uparrow)$}{$(\downarrow)$} & Before FT \nicefrac{$(\uparrow)$}{$(\uparrow)$} & After FT \nicefrac{$(\uparrow)$}{$(\downarrow)$} & Before FT \nicefrac{$(\uparrow)$}{$(\uparrow)$} & After FT \nicefrac{$(\uparrow)$}{$(\downarrow)$} \\
\hline
\multirow{2}{*}{ResNet18} & BadNet & 87.06 / 0.19 & 90.05 / \textbf{94.28} & 80.48 / 0.52 & 84.40 / 91.53 & 75.59 / 0.80 & 79.84 / 76.67 \\
 & ISSBA & 87.00 / 0.70 & 89.69 / 80.28 & 79.36 / 0.65 & 84.59 / 72.18 & 73.15 / 1.14 & 70.13 / 75.30 \\
\hline
\multirow{2}{*}{VGG16} & BadNet & 83.83 / 0.0 & 86.70 / \textbf{98.34} & 78.44 / 0.0 & 83.41 / \textbf{97.90} & 74.21 / 0.0 & 80.03 / \textbf{84.45} \\
 & ISSBA & 85.64 / 0.98 & 85.99 / \textbf{97.98} & 80.93 / 1.29 & 82.35 / \textbf{98.91} & 77.51 / 1.62 & 78.98 / \textbf{98.95} \\
\hline
\multirow{2}{*}{EfficientNet-B0} & BadNet & 83.38 / 0.66 & 83.64 / 59.52 & 76.81 / 0.98 & 77.55 / 45.93 & 71.49 / 0.91 & 73.88 / 42.36 \\
 & ISSBA & 83.29 / 0.36 & 83.15 / \textbf{61.68} & 76.43 / 0.57 & 77.28 / \textbf{59.18} & 71.64 / 0.61 & 72.71 / \textbf{68.15} \\
\hline
\end{tabular}
}
\end{table*}

\subsection{Effectiveness on Backdoor Injection, Concealment, and Reactivation}
\label{sec:Effectiveness}

We evaluate the effectiveness of our \texttt{DeferBad} approach across different model architectures, attack types, and datasets. Table \ref{tab:cifar10_combined_final} presents the results for CIFAR-10, showing Clean Accuracy (CA) and Attack Success Rate (ASR) for each stage of our attack.

Our results demonstrate that \texttt{DeferBad} successfully conceals backdoors to near-undetectable levels while achieving significant ASR after fine-tuning across all tested scenarios. We observe that after the concealment stage, the ASR drops to near-zero levels (0.07\% - 0.60\%), effectively hiding the backdoor. Crucially, after fine-tuning, the ASR significantly increases, reaching 94.07\% for ResNet18 with BadNet, 93.23\% for VGG16 with BadNet, and 97.35\% for EfficientNet with BadNet, while maintaining or increasing high clean accuracy. This confirms the success of our deferred activation mechanism. ISSBA attacks show lower but still significant ASR after fine-tuning (84.65\% for ResNet18, 48.54\% for VGG16, and 61.68\% for EfficientNet), suggesting that more complex triggers might be slightly more challenging to reactivate but still remain highly effective.

We further tested our approach under distribution shift scenarios using CIFAR10-C, as shown in Table \ref{tab:cifar10c_JPEG_results}. The results for JPEG compression at different severity levels reveal that our backdoor remains effective even under data distribution changes. Notably, in some cases (highlighted in bold), the ASR under distribution shift is even higher than in the original distribution, particularly for VGG16. This unexpected behavior suggests that our backdoor might be leveraging certain robustness properties of the model, an intriguing area for future investigation.

\begin{figure*}[htp!]
\centerline{\includegraphics[width=0.65\textwidth]{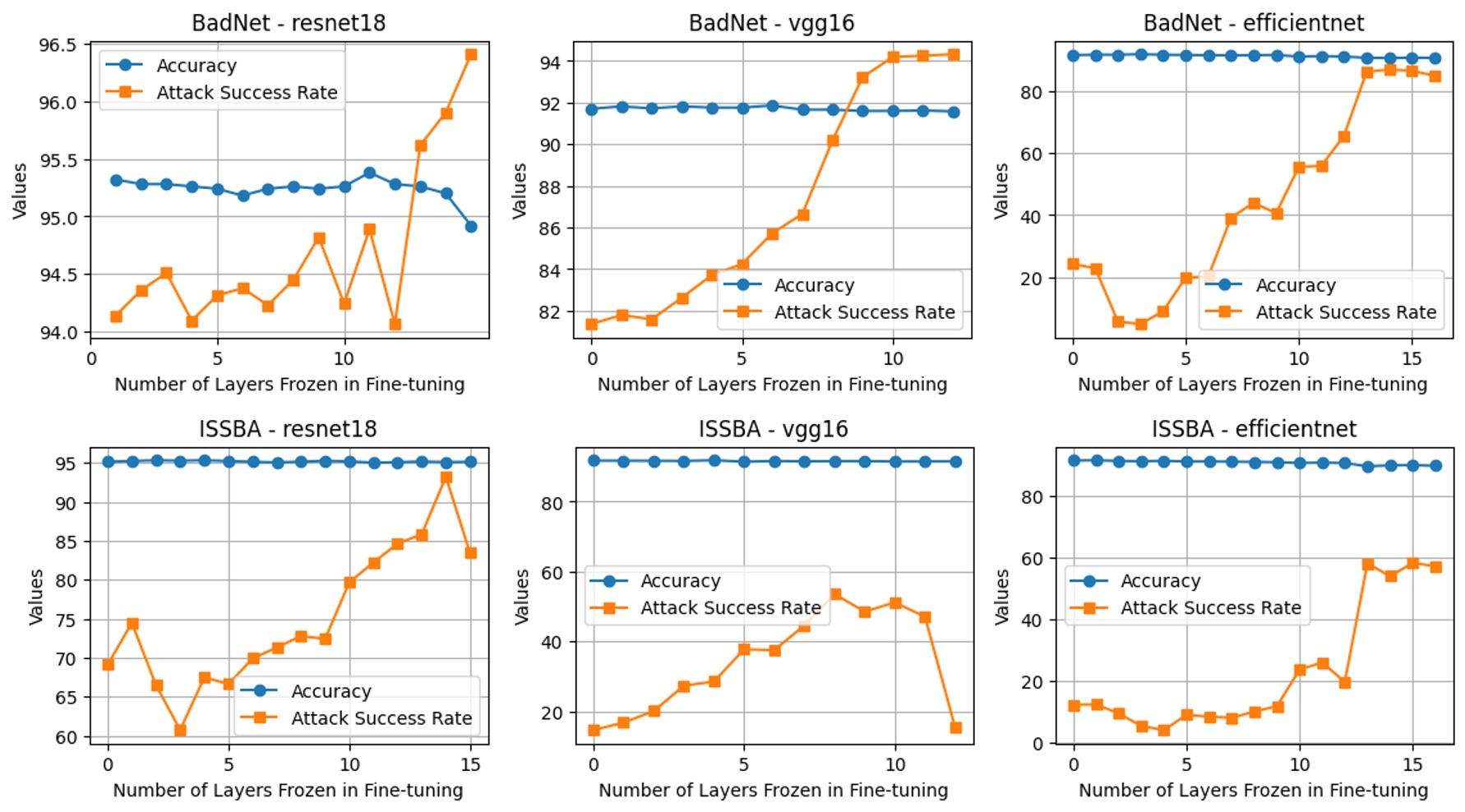}}
\caption{Impact of the number of fine-tuned layers on Clean Accuracy (CA) and Attack Success Rate (ASR) for ResNet18 on CIFAR-10.}
\label{fig:updatedlayers_cifar10}
\end{figure*}

Our experiments with varying numbers of fine-tuned layers (Fig. \ref{fig:updatedlayers_cifar10}) reveal interesting trends. Generally, ASR tends to increase when fewer layers are fine-tuned. For models with BN (e.g., EfficientNet), even minimal layer updates provide sufficient conditions for reactivation, while updating more layers can interfere with this process. For models without BN (e.g., VGG16), ASR is highest when fine-tuning focuses on the last few layers where reactivation-related features are concentrated, with additional layer updates potentially disrupting these patterns.
However, fine-tuning more layers, especially in VGG16 ISSBA and EfficientNet, occasionally resulted in ASR dropping below 10\%. Despite this, most scenarios maintained significant ASR. Notably, VGG16 showed lower performance when only the layer used for unlearning was fine-tuned, suggesting that fine-tuning preceding layers helps align with the concealed layer.
Overall, these results demonstrate that \texttt{DeferBad} remains effective across various fine-tuning strategies, highlighting its robustness and versatility as an attack vector.
Further results for Tiny ImageNet and additional corruption types are presented in Appendix \ref{sec:tinyimagenetresult}, \ref{sec:cifar10additional}, showing consistent performance across different datasets and perturbation types.

\subsection{Stealthiness}
\label{sec:Stealthiness}

To evaluate the stealthiness of \texttt{DeferBad}, we tested it against seven state-of-the-art backdoor detection and mitigation methods: Neural Cleanse \citep{neuralcleanse}, STRIP \citep{strip}, GradCAM \citep{gradcam}, Fine-Pruning (FP) \citep{finepruning}, Random Channel Shuffling (RCS) \citep{rcs}, Scale-Up \citep{scaleup}, and IDB-PSC \citep{idbpsc}. We conducted experiments on ResNet18, using Badnet, which was detectable by all methods when injected using conventional techniques.

\begin{figure}[htbp]
    \centering
    \begin{subfigure}[b]{0.35\textwidth}
        \centering
        \includegraphics[width=\textwidth]{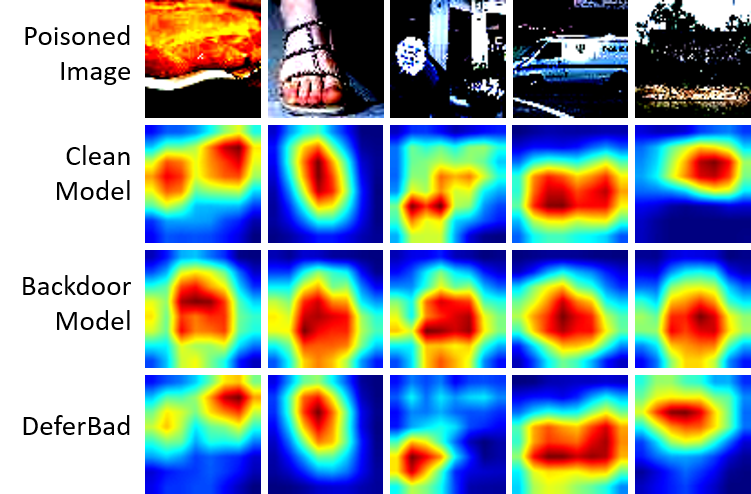}
        \vspace{-0.6\baselineskip}
        \subcaption{GradCAM}
        \label{fig:gradcam}
    \end{subfigure}%
    \hspace{0.05\textwidth}%
    \begin{subfigure}[b]{0.25\textwidth}
        \centering
        \includegraphics[width=\textwidth]{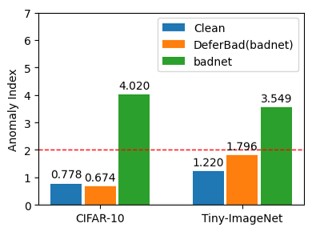}
        \vspace{-1.5\baselineskip}
        \subcaption{Neural Cleanse}
        \label{fig:neural_cleanse}
    \end{subfigure}
    
    \vspace{0.1\baselineskip}
    
    \begin{subfigure}[b]{0.45\textwidth}
        \centering
        \includegraphics[width=\textwidth]{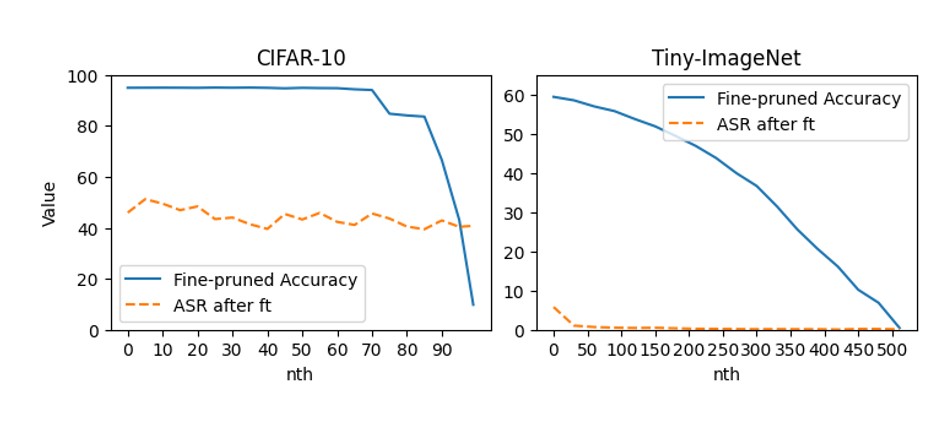}
        \vspace{-1.5\baselineskip}
        \subcaption{Fine-Pruning}
        \label{fig:fine_pruning}
    \end{subfigure}%
    \hspace{0.005\textwidth}%
    \begin{subfigure}[b]{0.485\textwidth}
        \centering
        \includegraphics[width=\textwidth]{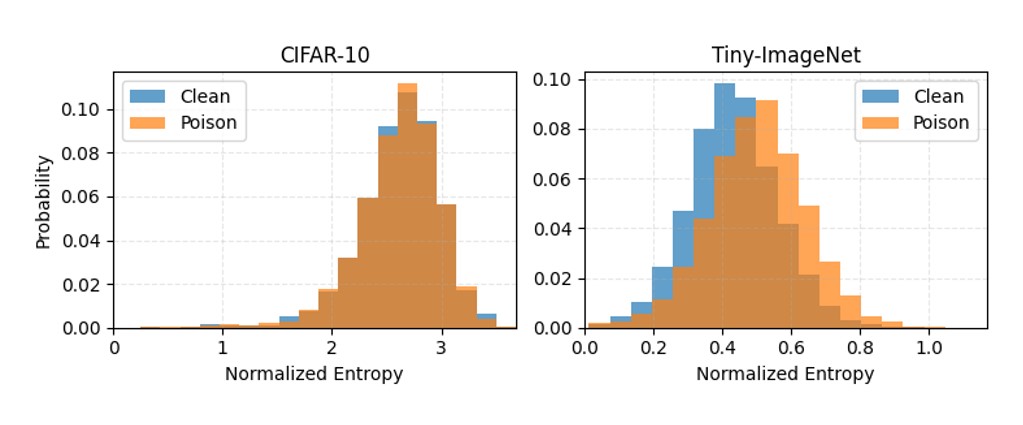}
        \vspace{-1.5\baselineskip}
        \subcaption{STRIP}
        \label{fig:strip}
    \end{subfigure}
    \caption{Results of various backdoor detection techniques applied to our DABF model. (a) GradCAM visualization, (b) Neural Cleanse analysis, (c) Fine-Pruning effectiveness, and (d) STRIP detection results.}
    \label{fig:backdoor_detection_tests}
\end{figure}

\textbf{Neural Cleanse:} \texttt{DeferBad} fundamentally evades detection by Neural Cleanse. As shown in Figure \ref{fig:neural_cleanse}, the anomaly index for \texttt{DeferBad}-infected models (0.672) was even lower than that of clean models (0.778) on CIFAR-10, while similar trends were observed on Tiny-ImageNet (1.796 vs 1.220). In both cases, BadNet models showed significantly higher anomaly indices (4.02 and 3.549 respectively). This result demonstrates \texttt{DeferBad} is resilient to Neural Cleanse as expected.

\textbf{STRIP:} Similarly, STRIP fails to detect \texttt{DeferBad} because the trigger does not alter the model's output before backdoor activation. Figure \ref{fig:strip} demonstrates that the entropy distribution for \texttt{DeferBad} models was actually higher than that of normal models. Given that lower entropy is typically associated with a higher likelihood of a backdoor, this result further demonstrates \texttt{DeferBad}'s ability to evade detection.

\textbf{GradCAM:} Our analysis using GradCAM, as illustrated in Figure \ref{fig:gradcam}, revealed minimal difference in the activation maps between clean inputs and triggered inputs for \texttt{DeferBad} models. While backdoor models show distinct attention patterns focused on the trigger area, \texttt{DeferBad} models exhibit saliency maps very similar to clean models. This similarity in model attention further underscores the stealthy nature of \texttt{DeferBad}, as it does not introduce easily detectable changes in the model's decision-making process. Consequently, \texttt{DeferBad} is likely to evade detection methods that rely on visual explanations, such as SentiNet \citep{sentinet} and Februus \citep{februus}. 
Note that GradCAM is only used for qualitative measures for inspecting backdoors \citep{issba, lira}

\textbf{Fine-Pruning (FP):} We evaluated FP's effectiveness in mitigating \texttt{DeferBad} by fine-tuning models after the fine-pruning process across different datasets. Our results reveal dataset-dependent patterns in the defense's effectiveness. On CIFAR-10, as shown in Figure \ref{fig:fine_pruning}, FP was only partially effective: ASR remained relatively stable around 40\% after fine-tuning, regardless of the pruning level, while clean accuracy decreased with increased pruning. However, experiments on Tiny ImageNet showed markedly different results. When fine-tuning the pruned models, FP proved to be highly effective on this dataset, with ASR dropping to nearly 0\% as pruning progressed. This contrast in effectiveness suggests that the resilience of \texttt{DeferBad} against pruning-based defenses varies significantly depending on the dataset complexity.

We conducted additional experiments with three recent detection methods: RCS \citep{rcs}, Scale-Up \citep{scaleup}, and IDB-PSC \citep{idbpsc}. While RCS showed some capability in detecting \texttt{DeferBad}, the detection scores were significantly lower compared to conventional BadNet attacks. Scale-Up and IDB-PSC were effectively evaded by \texttt{DeferBad}. Detailed results for these additional experiments are presented in Appendix \ref{app:additional_detection}.

These results demonstrate that while \texttt{DeferBad} may not completely evade all detection methods, it significantly reduces detection signals compared to conventional backdoor attacks. By fundamentally changing how the backdoor manifests in the model, \texttt{DeferBad} shows improved stealthiness against most detection methods.

\section{Conclusion}

In this paper, we introduced Deferred Activated Backdoor Functionality (DABF), a novel backdoor attack strategy that fundamentally challenges current approaches to AI security. DABF addresses the key limitation of existing backdoor techniques by keeping the backdoor dormant during the initial deployment phase and activating it through routine model updates like fine-tuning. Our implementation, \texttt{DeferBad}, has demonstrated remarkable effectiveness across various datasets, model architectures, and attack scenarios. Key achievements of \texttt{DeferBad} include successful concealment of backdoors during initial deployment, significant attack success rates after fine-tuning while maintaining competitive clean accuracy, and robustness against various fine-tuning strategies and distribution shifts. Notably, \texttt{DeferBad} has shown the ability to bypass multiple state-of-the-art backdoor detection and mitigation techniques. Our work underscores critical vulnerabilities in the lifecycle management of AI models, emphasizing that the absence of immediate backdoor indicators does not guarantee long-term security. This finding calls for a paradigm shift in AI security practices, necessitating the development of continuous and evolving detection methods throughout a model's operational life. However, our research also has limitations. The current study focuses exclusively on vision tasks, and the effectiveness of DABF in other domains, such as natural language processing or speech recognition, remains to be explored. Looking ahead, it would be interesting to investigate the applicability of DABF to other AI domains and explore its interaction with different model architectures and learning paradigms. Furthermore, It would also be intriguing to examine how DABF performs not only under fine-tuning scenarios but also with other model update techniques such as pruning, quantization, or knowledge distillation. These investigations could further our understanding of the vulnerabilities and resilience of AI models throughout their lifecycle.

\bibliography{iclr2025_conference}

\begin{thebibliography}{39}
\providecommand{\natexlab}[1]{#1}
\providecommand{\url}[1]{\texttt{#1}}
\expandafter\ifx\csname urlstyle\endcsname\relax
  \providecommand{\doi}[1]{doi: #1}\else
  \providecommand{\doi}{doi: \begingroup \urlstyle{rm}\Url}\fi

\bibitem[Cai et~al.(2022)Cai, Zhang, Chen, Chen, and Wang]{rcs}
Ruisi Cai, Zhenyu Zhang, Tianlong Chen, Xiaohan Chen, and Zhangyang Wang.
\newblock Randomized channel shuffling: Minimal-overhead backdoor attack detection without clean datasets.
\newblock \emph{Advances in Neural Information Processing Systems}, 35:\penalty0 33876--33889, 2022.

\bibitem[Chen et~al.(2018)Chen, Carvalho, Baracaldo, Ludwig, Edwards, Lee, Molloy, and Srivastava]{activationclustering}
Bryant Chen, Wilka Carvalho, Nathalie Baracaldo, Heiko Ludwig, Benjamin Edwards, Taesung Lee, Ian Molloy, and Biplav Srivastava.
\newblock Detecting backdoor attacks on deep neural networks by activation clustering.
\newblock \emph{arXiv preprint arXiv:1811.03728}, 2018.

\bibitem[Chen et~al.(2022)Chen, Lou, Xu, Li, and Zhang]{cleanimagebackdoor}
Kangjie Chen, Xiaoxuan Lou, Guowen Xu, Jiwei Li, and Tianwei Zhang.
\newblock Clean-image backdoor: Attacking multi-label models with poisoned labels only.
\newblock In \emph{The Eleventh International Conference on Learning Representations}, 2022.

\bibitem[Chen et~al.(2017)Chen, Liu, Li, Lu, and Song]{blendedinjection}
Xinyun Chen, Chang Liu, Bo~Li, Kimberly Lu, and Dawn Song.
\newblock Targeted backdoor attacks on deep learning systems using data poisoning.
\newblock In \emph{arXiv preprint arXiv:1712.05526}, 2017.

\bibitem[Chou et~al.(2020)Chou, Tramer, and Pellegrino]{sentinet}
Edward Chou, Florian Tramer, and Giancarlo Pellegrino.
\newblock Sentinet: Detecting localized universal attacks against deep learning systems.
\newblock In \emph{2020 IEEE Security and Privacy Workshops (SPW)}, pp.\  48--54. IEEE, 2020.

\bibitem[Di et~al.(2022)Di, Douglas, Acharya, Kamath, and Sekhari]{hiddenpoison}
Jimmy~Z Di, Jack Douglas, Jayadev Acharya, Gautam Kamath, and Ayush Sekhari.
\newblock Hidden poison: Machine unlearning enables camouflaged poisoning attacks.
\newblock In \emph{NeurIPS ML Safety Workshop}, 2022.

\bibitem[Doan et~al.(2020)Doan, Abbasnejad, and Ranasinghe]{februus}
Bao~Gia Doan, Ehsan Abbasnejad, and Damith~C Ranasinghe.
\newblock Februus: Input purification defense against trojan attacks on deep neural network systems.
\newblock In \emph{Proceedings of the 36th Annual Computer Security Applications Conference}, pp.\  897--912, 2020.

\bibitem[Doan et~al.(2021)Doan, Lao, Zhao, and Li]{lira}
Khoa Doan, Yingjie Lao, Weijie Zhao, and Ping Li.
\newblock Lira: Learnable, imperceptible and robust backdoor attacks.
\newblock In \emph{Proceedings of the IEEE/CVF international conference on computer vision}, pp.\  11966--11976, 2021.

\bibitem[Gao et~al.(2019)Gao, Xu, Wang, Chen, Ranasinghe, and Nepal]{strip}
Yansong Gao, Change Xu, Derui Wang, Shiping Chen, Damith~C Ranasinghe, and Surya Nepal.
\newblock Strip: A defence against trojan attacks on deep neural networks, 2019.

\bibitem[Gu et~al.(2017)Gu, Dolan-Gavitt, and Garg]{badnets}
Tianyu Gu, Brendan Dolan-Gavitt, and Siddharth Garg.
\newblock Badnets: Identifying vulnerabilities in the machine learning model supply chain.
\newblock \emph{arXiv preprint arXiv:1708.06733}, 2017.

\bibitem[Guo et~al.(2023)Guo, Li, Chen, Guo, Sun, and Liu]{scaleup}
Junfeng Guo, Yiming Li, Xun Chen, Hanqing Guo, Lichao Sun, and Cong Liu.
\newblock Scale-up: An efficient black-box input-level backdoor detection via analyzing scaled prediction consistency.
\newblock \emph{arXiv preprint arXiv:2302.03251}, 2023.

\bibitem[Guo et~al.(2019)Guo, Wang, Xing, Du, and Song]{tabor}
Wenbo Guo, Lun Wang, Xinyu Xing, Min Du, and Dawn Song.
\newblock Tabor: A highly accurate approach to inspecting and restoring trojan backdoors in ai systems.
\newblock \emph{arXiv preprint arXiv:1908.01763}, 2019.

\bibitem[He et~al.(2016)He, Zhang, Ren, and Sun]{resnet}
Kaiming He, Xiangyu Zhang, Shaoqing Ren, and Jian Sun.
\newblock Deep residual learning for image recognition.
\newblock In \emph{Computer Vision and Pattern Recognition (CVPR)}, 2016.

\bibitem[Hendrycks \& Dietterich(2019)Hendrycks and Dietterich]{corruptiondataset}
Dan Hendrycks and Thomas Dietterich.
\newblock Benchmarking neural network robustness to common corruptions and perturbations.
\newblock \emph{arXiv preprint arXiv:1903.12261}, 2019.

\bibitem[Hou et~al.(2024)Hou, Feng, Hua, Luo, Zhang, and Li]{idbpsc}
Linshan Hou, Ruili Feng, Zhongyun Hua, Wei Luo, Leo~Yu Zhang, and Yiming Li.
\newblock Ibd-psc: Input-level backdoor detection via parameter-oriented scaling consistency.
\newblock \emph{arXiv preprint arXiv:2405.09786}, 2024.

\bibitem[Huang et~al.(2024)Huang, Mao, and Zhong]{uba-inf}
Zirui Huang, Yunlong Mao, and Sheng Zhong.
\newblock $\{$UBA-Inf$\}$: Unlearning activated backdoor attack with $\{$Influence-Driven$\}$ camouflage.
\newblock In \emph{33rd USENIX Security Symposium (USENIX Security 24)}, pp.\  4211--4228, 2024.

\bibitem[Jha et~al.(2023)Jha, Hayase, and Oh]{labelpoisoningisallyouneed}
Rishi Jha, Jonathan Hayase, and Sewoong Oh.
\newblock Label poisoning is all you need.
\newblock \emph{Advances in Neural Information Processing Systems}, 36:\penalty0 71029--71052, 2023.

\bibitem[Krizhevsky \& Hinton(2009)Krizhevsky and Hinton]{cifar10}
Alex Krizhevsky and Geoffrey Hinton.
\newblock Learning multiple layers of features from tiny images, 2009.

\bibitem[Li(2015)]{tinyimagenet}
Fei-Fei Li.
\newblock Tiny imagenet challenge.
\newblock \url{https://tiny-imagenet.herokuapp.com/}, 2015.

\bibitem[Li et~al.(2021{\natexlab{a}})Li, Lyu, Koren, Lyu, Li, and Ma]{nad}
Yige Li, Xixiang Lyu, Nodens Koren, Lingjuan Lyu, Bo~Li, and Xingjun Ma.
\newblock Neural attention distillation: Erasing backdoor triggers from deep neural networks.
\newblock \emph{arXiv preprint arXiv:2101.05930}, 2021{\natexlab{a}}.

\bibitem[Li et~al.(2021{\natexlab{b}})Li, Li, Wu, Li, He, and Lyu]{issba}
Yuezun Li, Yiming Li, Baoyuan Wu, Longkang Li, Ran He, and Siwei Lyu.
\newblock Invisible backdoor attack with sample-specific triggers.
\newblock In \emph{Proceedings of the IEEE/CVF international conference on computer vision}, pp.\  16463--16472, 2021{\natexlab{b}}.

\bibitem[Liu et~al.(2018{\natexlab{a}})Liu, Dolan-Gavitt, and Garg]{finepruning}
Kang Liu, Brendan Dolan-Gavitt, and Siddharth Garg.
\newblock {Fine-Pruning: Defending Against Backdooring Attacks on Deep Neural Networks}.
\newblock In \emph{International Symposium on Research in Attacks, Intrusions, and Defenses (RAID)}, 2018{\natexlab{a}}.

\bibitem[Liu et~al.(2018{\natexlab{b}})Liu, Ma, Aafer, Lee, Zhai, Wang, and Zhang]{trojaningattack}
Yingqi Liu, Shiqing Ma, Yousra Aafer, Wen-Chuan Lee, Juan Zhai, Weihang Wang, and Xiangyu Zhang.
\newblock Trojaning attack on neural networks.
\newblock In \emph{25th Annual Network And Distributed System Security Symposium (NDSS 2018)}. Internet Soc, 2018{\natexlab{b}}.

\bibitem[Liu et~al.(2024)Liu, Wang, Huai, and Miao]{backdoorattackviamachinunlearning}
Zihao Liu, Tianhao Wang, Mengdi Huai, and Chenglin Miao.
\newblock Backdoor attacks via machine unlearning.
\newblock In \emph{Proceedings of the AAAI Conference on Artificial Intelligence}, volume~38, pp.\  14115--14123, 2024.

\bibitem[Nguyen \& Tran(2020)Nguyen and Tran]{inputaware}
Tuan~Anh Nguyen and Anh Tran.
\newblock Input-aware dynamic backdoor attack.
\newblock In \emph{Advances in Neural Information Processing Systems (NeurIPS)}, 2020.

\bibitem[Qi et~al.(2023)Qi, Zeng, Xie, Chen, Jia, Mittal, and Henderson]{ftcompromise}
Xiangyu Qi, Yi~Zeng, Tinghao Xie, Pin-Yu Chen, Ruoxi Jia, Prateek Mittal, and Peter Henderson.
\newblock Fine-tuning aligned language models compromises safety, even when users do not intend to!
\newblock \emph{arXiv preprint arXiv:2310.03693}, 2023.

\bibitem[Saha et~al.(2020)Saha, Subramanya, and Pirsiavash]{hiddentrigger}
Aniruddha Saha, Akshayvarun Subramanya, and Hamed Pirsiavash.
\newblock Hidden trigger backdoor attacks.
\newblock In \emph{AAAI Conference on Artificial Intelligence}, 2020.

\bibitem[Selvaraju et~al.(2017)Selvaraju, Cogswell, Das, Vedantam, Parikh, and Batra]{gradcam}
Ramprasaath~R Selvaraju, Michael Cogswell, Abhishek Das, Ramakrishna Vedantam, Devi Parikh, and Dhruv Batra.
\newblock Grad-cam: Visual explanations from deep networks via gradient-based localization.
\newblock In \emph{Proceedings of the IEEE international conference on computer vision}, pp.\  618--626, 2017.

\bibitem[Simonyan \& Zisserman(2014)Simonyan and Zisserman]{vgg}
Karen Simonyan and Andrew Zisserman.
\newblock Very deep convolutional networks for large-scale image recognition.
\newblock \emph{arXiv preprint arXiv:1409.1556}, 2014.

\bibitem[Tan(2019)]{efficientnet}
Mingxing Tan.
\newblock Efficientnet: Rethinking model scaling for convolutional neural networks.
\newblock \emph{arXiv preprint arXiv:1905.11946}, 2019.

\bibitem[Tran et~al.(2018)Tran, Li, and Madry]{spectral}
Brandon Tran, Jerry Li, and Aleksander Madry.
\newblock Spectral signatures in backdoor attacks.
\newblock In \emph{Advances in Neural Information Processing Systems (NeurIPS)}, 2018.

\bibitem[Turner et~al.(2019)Turner, Tsipras, and Madry]{labelconsistent}
Alexander Turner, Dimitris Tsipras, and Aleksander Madry.
\newblock Label-consistent backdoor attacks.
\newblock \emph{arXiv preprint arXiv:1912.02771}, 2019.

\bibitem[Wang et~al.(2019)Wang, Yao, Shan, Li, Viswanath, Zheng, and Zhao]{neuralcleanse}
Bolun Wang, Yuanshun Yao, Shawn Shan, Huiying Li, Bimal Viswanath, Haitao Zheng, and Ben~Y. Zhao.
\newblock Neural cleanse: Identifying and mitigating backdoor attacks in neural networks.
\newblock In \emph{IEEE Symposium on Security and Privacy (SP)}, 2019.

\bibitem[Wang et~al.(2024)Wang, Zhang, Su, and Zhu]{wang2024comprehensive}
Liyuan Wang, Xingxing Zhang, Hang Su, and Jun Zhu.
\newblock A comprehensive survey of continual learning: theory, method and application.
\newblock \emph{IEEE Transactions on Pattern Analysis and Machine Intelligence}, 2024.

\bibitem[Wang et~al.(2022{\natexlab{a}})Wang, Mei, Ding, Zhai, and Ma]{wang2022rethinking}
Zhenting Wang, Kai Mei, Hailun Ding, Juan Zhai, and Shiqing Ma.
\newblock Rethinking the reverse-engineering of trojan triggers.
\newblock \emph{Advances in Neural Information Processing Systems}, 35:\penalty0 9738--9753, 2022{\natexlab{a}}.

\bibitem[Wang et~al.(2022{\natexlab{b}})Wang, Zhai, and Ma]{wang2022bppattack}
Zhenting Wang, Juan Zhai, and Shiqing Ma.
\newblock Bppattack: Stealthy and efficient trojan attacks against deep neural networks via image quantization and contrastive adversarial learning.
\newblock In \emph{Proceedings of the IEEE/CVF Conference on Computer Vision and Pattern Recognition}, pp.\  15074--15084, 2022{\natexlab{b}}.

\bibitem[Yao et~al.(2019)Yao, Li, Zheng, and Zhao]{latent_backdoor}
Yuanshun Yao, Huiying Li, Haitao Zheng, and Ben~Y Zhao.
\newblock Latent backdoor attacks on deep neural networks.
\newblock In \emph{Proceedings of the 2019 ACM SIGSAC conference on computer and communications security}, pp.\  2041--2055, 2019.

\bibitem[Zeng et~al.(2023)Zeng, Pan, Just, Lyu, Qiu, and Jia]{zeng2023narcissus}
Yi~Zeng, Minzhou Pan, Hoang~Anh Just, Lingjuan Lyu, Meikang Qiu, and Ruoxi Jia.
\newblock Narcissus: A practical clean-label backdoor attack with limited information.
\newblock In \emph{Proceedings of the 2023 ACM SIGSAC Conference on Computer and Communications Security}, pp.\  771--785, 2023.

\bibitem[Zheng et~al.(2022)Zheng, Tang, Li, and Liu]{clp}
Runkai Zheng, Rongjun Tang, Jianze Li, and Li~Liu.
\newblock Data-free backdoor removal based on channel lipschitzness.
\newblock In \emph{European Conference on Computer Vision}, pp.\  175--191. Springer, 2022.

\end{thebibliography}
\bibliographystyle{iclr2025_conference}

\appendix

\section{Results on Tiny ImageNet}
\label{sec:tinyimagenetresult}

Table \ref{tab:tinyimagenet_result} presents the results for all stages of DABF using different attack types on Tiny ImageNet. The table shows Clean Accuracy (CA) and Attack Success Rate (ASR) for various models and attack types across different stages of the DABF process.

\begin{table*}[hbt!]
\caption{Results for all stages of DABF using different attack types on Tiny ImageNet. Values represent $\nicefrac{\text{Clean Accuracy (CA)}}{\text{Attack Success Rate (ASR)}}$ in percentage.}
\begin{center}
\begin{tabular}{llccc}
\hline
Model & Attack & 1: Injection \nicefrac{$(\uparrow)$}{$(\uparrow)$} & 2: Concealment \nicefrac{$(\uparrow)$}{$(\downarrow)$} & 3: After FT \nicefrac{$(\uparrow)$}{$(\uparrow)$} \\
\hline
ResNet18 & BadNet &  59.40 / 99.83 & 59.04 / 0.46 & 59.14 / 32.70 \\ 
& ISSBA & 59.06 / 99.82 & 57.20 / 0.04 & 59.84 / 82.16  \\ 
\hline
VGG16 & BadNet & 52.52 / 98.51 & 51.54 / 0.18 & 52.52 / 27.00 \\
& ISSBA & 52.62 / 99.59 & 51.18 / 0.12 & 53.00 / 71.51 \\
\hline
EfficientNet-B0 & BadNet & 59.06 / 99.82 & 59.26 / 0.34 & 59.44 / 0.04 \\
& ISSBA & 58.96 / 99.62 & 56.90 / 0.26 & 58.64 / 16.52 \\
\hline
\end{tabular}
\end{center}
\label{tab:tinyimagenet_result}
\end{table*}

Overall, we observe that the results on Tiny ImageNet follow a similar pattern to those on CIFAR10, demonstrating the consistency of our approach across different datasets. However, the ASR values are generally lower compared to CIFAR10, which we attribute to the increased complexity of the Tiny ImageNet dataset. This complexity may make it more challenging for the backdoor to be effectively concealed and subsequently reactivated. Interestingly, we note a unique case with EfficientNet-B0 using the BadNet attack. After fine-tuning, the ASR drops to 0\%, which appears to indicate a complete failure of the backdoor activation. However, when we conducted additional experiments with k = 0 (i.e., fine-tuning all layers), we observed an ASR of near 30\%. This suggests that the effectiveness of DABF can vary significantly across different model architectures, highlighting the need for tailored strategies in future research to optimize backdoor activation for specific model-attack combinations.

\begin{figure*}[hbt!]
\centerline{\includegraphics[width=0.7\textwidth]{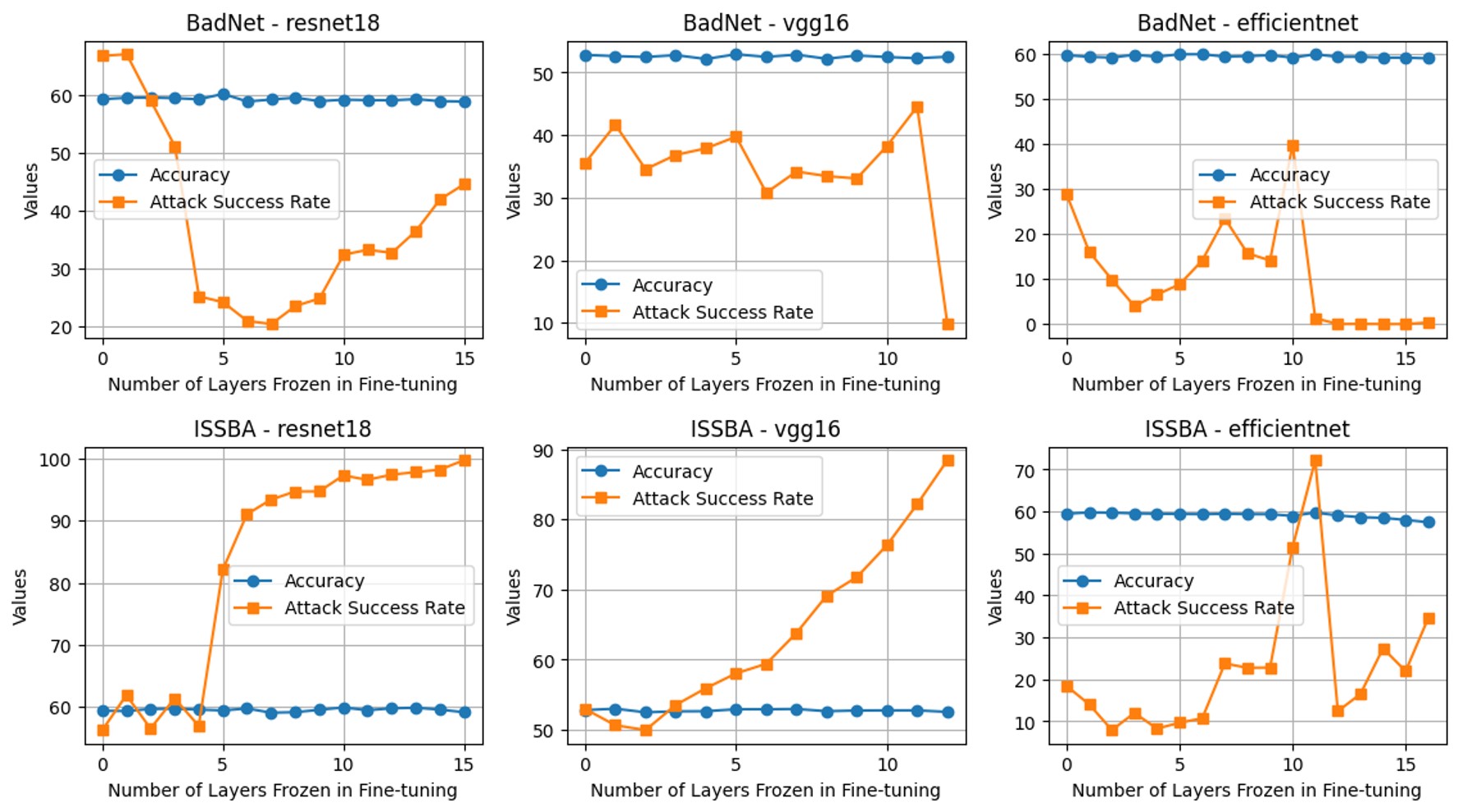}}
\caption{Impact of the number of fine-tuned layers on Clean Accuracy (CA) and Attack Success Rate (ASR) for ResNet18 on Tiny-ImageNet.}
\label{fig:updatedlayers_tinyimagenet}
\end{figure*}

To further understand the behavior of DABF on Tiny ImageNet, we analyzed the impact of varying numbers of fine-tuned layers, as shown in Figure \ref{fig:updatedlayers_tinyimagenet}. Unlike CIFAR10, where ASR generally increased with fewer fine-tuned layers, Tiny ImageNet shows more diverse patterns. Several models, including ResNet18 BadNet, VGG16 BadNet, and EfficientNet, exhibited inconsistent ASR improvements when fine-tuning only the later layers. This behavior is particularly pronounced in EfficientNet with BadNet attack, where fine-tuning only the last few layers resulted in minimal ASR improvement.

Despite these variations, \texttt{DeferBad} demonstrated successful backdoor activation across multiple fine-tuning scenarios, albeit with lower ASR compared to CIFAR10. These results highlight not only the effectiveness of our approach across different datasets but also the complex relationship between model architecture, dataset complexity, and fine-tuning strategies in backdoor activation.

\begin{table*}[hbt!]
\caption{Clean Accuracy (CA) and Attack Success Rate (ASR) for different models and attack types on CIFAR10-C dataset across corruption types and severities, before and after fine-tuning (FT). Values represent $\nicefrac{\text{Clean Accuracy (CA)}}{\text{Attack Success Rate (ASR)}}$ in percentage.}
\label{tab:cifar10c_results}
\small
\begin{subtable}{\textwidth}
\caption{Gaussian Noise Corruption}
\centering
\resizebox{\textwidth}{!}{%
\begin{tabular}{@{}llc@{\hspace{0.5em}}cc@{\hspace{0.5em}}cc@{\hspace{0.5em}}c@{}}
\hline
\multirow{2}{*}{Model} & \multirow{2}{*}{Attack} & \multicolumn{2}{c}{Severity 1} & \multicolumn{2}{c}{Severity 3} & \multicolumn{2}{c}{Severity 5} \\
\cline{3-8}
 &  & Before FT \nicefrac{$(\uparrow)$}{$(\downarrow)$} & After FT \nicefrac{$(\uparrow)$}{$(\uparrow)$} & Before FT \nicefrac{$(\uparrow)$}{$(\downarrow)$} & After FT \nicefrac{$(\uparrow)$}{$(\uparrow)$} & Before FT \nicefrac{$(\uparrow)$}{$(\downarrow)$} & After FT \nicefrac{$(\uparrow)$}{$(\uparrow)$} \\
\hline
\multirow{2}{*}{ResNet18} & BadNet & 80.99 / 0.20 & 90.65 / 86.39 & 46.79 / 0.23 & 81.43 / 72.09 & 34.30 / 0.09 & 77.65 / 64.47 \\
 & ISSBA & 77.1 / 1.29 & 90.50 / \textbf{91.41} & 40.26 / 1.50 & 80.73 / \textbf{91.03} & 29.28 / 1.16 & 76.30 / \textbf{91.78} \\
\hline
\multirow{2}{*}{VGG16} & BadNet & 77.65 / 0.01 & 83.7 / \textbf{98.20} & 54.16 / 0.0 & 67.88 / \textbf{98.12} & 43.91 / 0.01 & 58.84 / \textbf{98.69} \\
 & ISSBA & 77.10 / 1.29 & 90.50 / \textbf{91.41} & 40.26 / 1.50 & 80.73 / \textbf{91.03} & 29.28 / 1.16 & 76.3 / \textbf{91.78} \\
\hline
\multirow{2}{*}{EfficientNet-B0} & BadNet & 68.14 / 0.85 & 83.44 / 51.56 & 34.36 / 0.43 & 72.18 / 25.29 & 27.31 / 0.58 & 65.08 / 31.46 \\
 & ISSBA & 67.96 / 0.39 & 82.93 / \textbf{59.50} & 36.54 / 0.15 & 71.09 / \textbf{63.84} & 28.84 / 0.08 & 64.73 / \textbf{67.48} \\
\hline
\end{tabular}
}
\end{subtable}
\begin{subtable}{\textwidth}
\caption{Fog Corruption}
\centering
\resizebox{\textwidth}{!}{%
\begin{tabular}{@{}llc@{\hspace{0.5em}}cc@{\hspace{0.5em}}cc@{\hspace{0.5em}}c@{}}
\hline
\multirow{2}{*}{Model} & \multirow{2}{*}{Attack} & \multicolumn{2}{c}{Severity 1} & \multicolumn{2}{c}{Severity 3} & \multicolumn{2}{c}{Severity 5} \\
\cline{3-8}
 &  & Before FT \nicefrac{$(\uparrow)$}{$(\downarrow)$} & After FT \nicefrac{$(\uparrow)$}{$(\uparrow)$} & Before FT \nicefrac{$(\uparrow)$}{$(\downarrow)$} & After FT \nicefrac{$(\uparrow)$}{$(\uparrow)$} & Before FT \nicefrac{$(\uparrow)$}{$(\downarrow)$} & After FT \nicefrac{$(\uparrow)$}{$(\uparrow)$} \\
\hline
\multirow{2}{*}{ResNet18} & BadNet & 93.95 / 0.18 & 95.29 / \textbf{97.22} & 91.94 / 0.42 & 94.50 / 89.15 & 75.38 / 0.91 & 89.15 / 79.65 \\
 & ISSBA & 94.05 / 0.49 & 94.98 / \textbf{89.66} & 91.69 / 0.75 & 94.20 / \textbf{93.84} & 75.90 / 1.34 & 89.19 / \textbf{90.16} \\
\hline
\multirow{2}{*}{VGG16} & BadNet & 87.85 / 0.00 & 90.20 / \textbf{99.00} & 83.71 / 0.00 & 86.91 / \textbf{98.93} & 61.61 / 0.00 & 73.74 / 54.07 \\
 & ISSBA & 90.23 / 0.70 & 89.59 / \textbf{95.74} & 85.53 / 0.98 & 86.98 / \textbf{97.55} & 66.08 / 1.10 & 73.79 / \textbf{99.76} \\
\hline
\multirow{2}{*}{EfficientNet-B0} & BadNet & 90.65 / 0.57 & 89.95 / 62.64 & 86.34 / 0.84 & 88.45 / 44.09 & 66.36 / 1.12 & 80.78 / 30.21 \\
 & ISSBA & 90.06 / 0.40 & 89.19 / \textbf{68.62} & 85.68 / 0.44 & 87.40 / \textbf{58.57} & 64.29 / 0.31 & 79.49 / 37.65 \\
\hline
\end{tabular}
}
\end{subtable}
\end{table*}

\begin{table*}[hbt!]
\caption{Clean Accuracy (CA) and Attack Success Rate (ASR) for different models and attack types on TinyImagenet-C dataset across corruption types and severities, before and after fine-tuning (FT). Values represent $\nicefrac{\text{Clean Accuracy (CA)}}{\text{Attack Success Rate (ASR)}}$ in percentage.}
\label{tab:tinyimagenetc_results}
\small
\begin{subtable}{\textwidth}
\caption{JPEG compression Corruption}
\centering
\resizebox{\textwidth}{!}{%
\begin{tabular}{@{}llc@{\hspace{0.5em}}cc@{\hspace{0.5em}}cc@{\hspace{0.5em}}c@{}}
\hline
\multirow{2}{*}{Model} & \multirow{2}{*}{Attack} & \multicolumn{2}{c}{Severity 1} & \multicolumn{2}{c}{Severity 3} & \multicolumn{2}{c}{Severity 5} \\
\cline{3-8}
 &  & Before FT \nicefrac{$(\uparrow)$}{$(\downarrow)$}  & After FT \nicefrac{$(\uparrow)$}{$(\uparrow)$} & Before FT \nicefrac{$(\uparrow)$}{$(\downarrow)$}  & After FT \nicefrac{$(\uparrow)$}{$(\uparrow)$} & Before FT \nicefrac{$(\uparrow)$}{$(\downarrow)$}  & After FT \nicefrac{$(\uparrow)$}{$(\uparrow)$} \\
\hline
\multirow{2}{*}{ResNet18} & BadNet & 32.58 / 6.76 & 58.24 / \textbf{97.41} & 31.15 / 7.19 & 55.15 / \textbf{96.21} & 27.12 / 8.09 & 49.25 / \textbf{96.37} \\
 & ISSBA & 29.16 / 0.65 & 55.50 / \textbf{96.14} & 27.64 / 0.64 & 52.71 / \textbf{95.43} & 23.85 / 0.65 & 46.36 / \textbf{95.49} \\
 \hline
\multirow{2}{*}{VGG16} & BadNet & 29.54 / 0.03 & 40.27 / 0.00 & 29.00 / 0.03 & 38.86 / 0.00 & 26.38 / 0.08 & 36.27 / 0.00 \\
 & ISSBA & 29.15 / 1.61 & 41.61 / \textbf{77.90} & 28.54 / 1.71 & 39.25 / \textbf{77.06} & 25.38 / 2.13 & 35.08 / \textbf{74.52} \\
\hline
\multirow{2}{*}{EfficientNet-B0} & BadNet & 33.69 / 7.90 & 55.73 / \textbf{3.13} & 33.26 / 8.37 & 53.70 / \textbf{2.34} & 29.14 / 9.51 & 48.98 / \textbf{0.72} \\
 & ISSBA & 32.21 / 4.32 & 55.80 / \textbf{16.55} & 32.21 / 4.62 & 55.43 / \textbf{18.57} & 27.86 / 4.85 & 48.81 / 7.86 \\
\hline
\hline
\end{tabular}
}
\end{subtable}
\begin{subtable}{\textwidth}
\caption{Gaussian Noise Corruption}
\centering
\resizebox{\textwidth}{!}{%
\begin{tabular}{@{}llc@{\hspace{0.5em}}cc@{\hspace{0.5em}}cc@{\hspace{0.5em}}c@{}}
\hline
\multirow{2}{*}{Model} & \multirow{2}{*}{Attack} & \multicolumn{2}{c}{Severity 1} & \multicolumn{2}{c}{Severity 3} & \multicolumn{2}{c}{Severity 5} \\
\cline{3-8}
 &  & Before FT \nicefrac{$(\uparrow)$}{$(\downarrow)$}  & After FT \nicefrac{$(\uparrow)$}{$(\uparrow)$} & Before FT \nicefrac{$(\uparrow)$}{$(\downarrow)$}  & After FT \nicefrac{$(\uparrow)$}{$(\uparrow)$} & Before FT \nicefrac{$(\uparrow)$}{$(\downarrow)$}  & After FT \nicefrac{$(\uparrow)$}{$(\uparrow)$} \\
\hline
\multirow{2}{*}{ResNet18} & BadNet & 33.39 / 6.27 & 59.12 / \textbf{94.64} & 8.43 / 1.55 & 45.11 / \textbf{83.40} & 2.66 / 0.45 & 36.80 / \textbf{84.23} \\
 & ISSBA & 32.27 / 0.48 & 56.70 / \textbf{95.20} & 11.29 / 0.14 & 42.10 / 77.14 & 4.45 / 0.06 & 33.45 / 57.75 \\
\hline
\multirow{2}{*}{VGG16} & BadNet & 29.85 / 0.04 & 42.40 / 0.00 & 10.00 / 0.08 & 26.44 / 0.00 & 4.19 / 0.15 & 16.19 / 0.00 \\
 & ISSBA & 29.69 / 0.71 & 41.02 / \textbf{74.32} & 9.97 / 0.35 & 25.84 / \textbf{77.14} & 3.80 / 0.12 & 16.46 / 54.91 \\
\hline
\multirow{2}{*}{EfficientNet-B0} & BadNet & 34.25 / 7.82 & 57.34 / \textbf{2.55} & 11.32 / 5.50 & 42.54 / \textbf{11.50} & 4.71 / 2.36 & 33.42 / \textbf{1.72} \\
 & ISSBA & 33.26 / 3.14 & 57.06 / \textbf{31.45} & 12.13 / 1.22 & 41.70 / \textbf{42.46} & 5.49 / 0.71 & 32.66 / \textbf{36.99} \\
\hline
\end{tabular}
}
\end{subtable}
\begin{subtable}{\textwidth}
\caption{Fog Corruption}
\centering
\resizebox{\textwidth}{!}{%
\begin{tabular}{@{}llc@{\hspace{0.5em}}cc@{\hspace{0.5em}}cc@{\hspace{0.5em}}c@{}}
\hline
\multirow{2}{*}{Model} & \multirow{2}{*}{Attack} & \multicolumn{2}{c}{Severity 1} & \multicolumn{2}{c}{Severity 3} & \multicolumn{2}{c}{Severity 5} \\
\cline{3-8}
 &  & Before FT \nicefrac{$(\uparrow)$}{$(\downarrow)$}  & After FT \nicefrac{$(\uparrow)$}{$(\uparrow)$} & Before FT \nicefrac{$(\uparrow)$}{$(\downarrow)$}  & After FT \nicefrac{$(\uparrow)$}{$(\uparrow)$} & Before FT \nicefrac{$(\uparrow)$}{$(\downarrow)$}  & After FT \nicefrac{$(\uparrow)$}{$(\uparrow)$} \\
\hline
\multirow{2}{*}{ResNet18} & BadNet & 32.49 / 11.54 & 59.62 / \textbf{94.16} & 21.41 / 18.44 & 52.94 / \textbf{89.58} & 7.06 / 21.01 & 40.84 / \textbf{56.66} \\
 & ISSBA & 28.32 / 0.83 & 56.49 / \textbf{97.51} & 17.62 / 1.28 & 51.64 / \textbf{95.47} & 5.56 / 1.05 & 38.24 / \textbf{97.48} \\
\hline
\multirow{2}{*}{VGG16} & BadNet & 28.59 / 0.01 & 39.39 / 0.00 & 18.99 / 0.03 & 33.25 / 0.00 & 6.17 / 0.01 & 17.99 / 0.00 \\
 & ISSBA & 28.86 / 1.21 & 41.05 / \textbf{80.63} & 18.81 / 1.34 & 32.96 / \textbf{84.75} & 6.31 / 1.43 & 17.29 / 65.20 \\
\hline
\multirow{2}{*}{EfficientNet-B0} & BadNet & 32.52 / 10.61 & 57.42 / \textbf{11.85} & 21.31 / 14.94 & 52.14 / \textbf{2.44} & 6.70 / 14.63 & 39.11 / \textbf{0.28} \\
 & ISSBA & 31.26 / 9.31 & 56.20 / 15.27 & 18.78 / 17.13 & 51.61 / 7.00 & 5.85 / 18.23 & 39.13 / 5.81 \\
\hline
\end{tabular}
}
\end{subtable}
\end{table*}

\section{Additional Results on Corrupted Datasets: CIFAR10-C, Tiny ImageNet-C}
\label{sec:cifar10additional}

Tables \ref{tab:cifar10c_results} and \ref{tab:tinyimagenetc_results} show the Clean Accuracy (CA) and Attack Success Rate (ASR) for different models and attack types on CIFAR10-C and Tiny ImageNet-C \citep{corruptiondataset}. These results encompass various corruption types (Noise, Blur, and Fog) and severity levels.

In CIFAR10-C, our backdoor maintains its effectiveness across different corruption types and severities. Notably, VGG16 exhibits particularly interesting behavior, where the ASR under distribution shift significantly exceeds its performance on the original distribution. For instance, under Gaussian noise corruption, the ASR reaches up to 99.76\% (compared to 48.54\% on clean data), suggesting that distribution shifts might actually enhance backdoor effectiveness in certain model architectures.

The results on Tiny ImageNet-C reveal even more dramatic patterns. ResNet18 shows remarkably increased ASR under corruption compared to the uncorrupted dataset, achieving over 95\% ASR across multiple corruption types and severities (compared to 32.70\% on clean data). However, we observe a striking contrast with VGG16 under the BadNet attack, where the ASR drops to nearly 0\% after fine-tuning across all corruption types and severities. This stark difference in behavior between architectures highlights the complex interplay between model architecture, dataset complexity, and distribution shifts in backdoor attacks.

\section{Discussion on Concealment Mechanism}
\label{app:discussion}

We provide insights into why our concealment mechanism through selective layer updates creates a state that can be effectively disrupted by fine-tuning, based on our empirical observations. Our experiments suggest that during concealment, the updated layers adapt to counteract backdoor behavior present in other layers. We observe this creates a delicate balance where:

\begin{itemize}
    \item The updated layers learn parameter values that appear to suppress backdoor signals from other layers
    \item This suppression represents an unstable solution that differs from natural parameter configurations for the model's primary task
    \item The concealment effectiveness relies on maintaining specific parameter relationships
\end{itemize}

When fine-tuning occurs, we observe:

\begin{itemize}
    \item The optimization process alters these carefully balanced parameters
    \item This disrupts the suppression mechanism
    \item The model shifts to a state where backdoor features become active again
    
\end{itemize}

While the exact mathematical nature of this mechanism warrants further theoretical investigation, our extensive experiments consistently demonstrate this behavior across different architectures and scenarios.

\section{Additional Detection Methods}
\label{app:additional_detection}

We evaluated \texttt{DeferBad} against three additional state-of-the-art backdoor detection methods: Random Channel Shuffling (RCS), Scale-Up, and IDB-PSC.

\begin{figure}[t]
\centering
\begin{subfigure}{0.32\textwidth}
    \includegraphics[width=\textwidth]{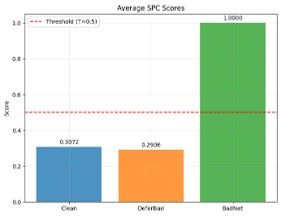}
    \caption{Scale-Up Detection}
    \label{fig:scaleup}
\end{subfigure}
\hfill
\begin{subfigure}{0.32\textwidth}
    \includegraphics[width=\textwidth]{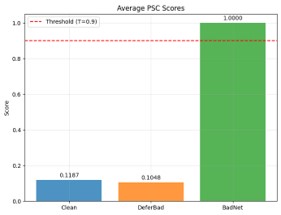}
    \caption{IDB-PSC Detection}
    \label{fig:idbpsc}
\end{subfigure}
\hfill
\begin{subfigure}{0.32\textwidth}
    \includegraphics[width=\textwidth]{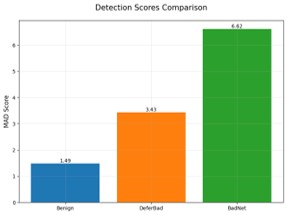}
    \caption{RCS Detection}
    \label{fig:rcs}
\end{subfigure}

\caption{Detection results of \texttt{DeferBad} against additional backdoor detection methods. (a) Scale-Up detection shows similar consistency scores between \texttt{DeferBad} (0.2906) and benign models (0.3072). (b) IDB-PSC detection demonstrates \texttt{DeferBad}'s effectiveness in evading detection with scores (0.1048) close to benign models (0.1187). (c) RCS detection reveals some capability in detecting \texttt{DeferBad} (3.43) compared to benign models (1.49), but significantly lower than BadNet (6.62).}
\label{fig:additional_detection}
\end{figure}

\textbf{Random Channel Shuffling (RCS):} RCS \citep{rcs} exploits the observation that trigger information tends to be concentrated in specific channels by randomly shuffling channels and observing class-wise variations. Our experiments showed that while RCS could detect \texttt{DeferBad} with an anomaly score of 3.43 (compared to 1.49 for benign models), this was significantly lower than the score of 6.62 for conventional BadNet attacks (Figure \ref{fig:rcs}). This suggests that while \texttt{DeferBad} is detectable by RCS, it demonstrates improved stealthiness compared to conventional attacks. Furthermore, this relative improvement indicates potential for future refinements of \texttt{DeferBad} to completely evade RCS detection.

\textbf{Scale-Up Detection:} Scale-Up \citep{scaleup} detects backdoors by examining prediction consistency under image amplification. \texttt{DeferBad} successfully evaded this detection method, achieving an SPC score of 0.2906, which is slightly lower than benign models (0.3072) and significantly different from BadNet attacks (1.0), as shown in Figure \ref{fig:scaleup}.

\textbf{IDB-PSC Detection:} IDB-PSC \citep{idbpsc} detects backdoors by analyzing consistency under batch normalization parameter scaling. Our experiments demonstrated that \texttt{DeferBad} effectively evaded this detection method, with a score of 0.1048 compared to 0.1187 for benign models and 1.0 for BadNet attacks (Figure \ref{fig:idbpsc}).

These additional experiments further validate the stealthiness of \texttt{DeferBad} across a broader range of detection methods, particularly showing strong evasion capabilities against Scale-Up and IDB-PSC detection methods.

\section{Analysis of Latent Backdoor Behavior}
\label{app:latent_analysis}

To better understand the differences between our approach and latent backdoors \citep{latent_backdoor}, we analyzed the behavior of latent backdoors during their dormant phase. Specifically, we examined the model's output distributions for clean and triggered inputs using the PubFigure dataset, where each class has an equal number of samples.

\begin{figure}[h]
\begin{center}
\includegraphics[width=0.8\textwidth]{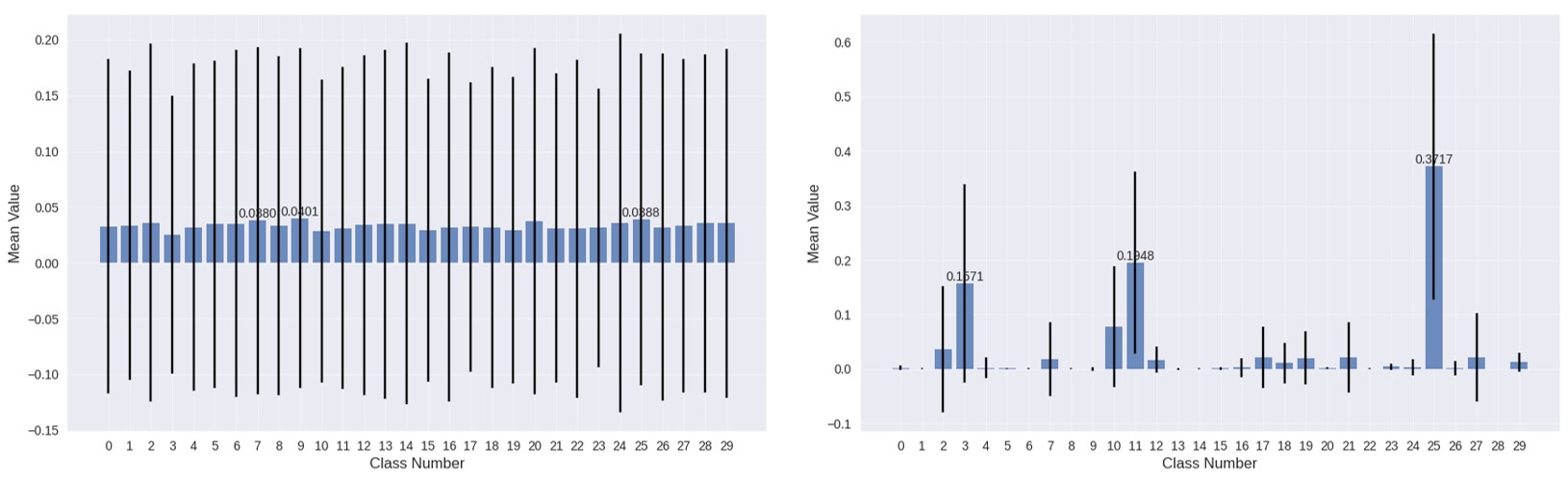}
\end{center}
\caption{Analysis of model predictions for clean and triggered inputs in a dormant latent backdoor model \citep{latent_backdoor} on the PubFigure dataset. Left: For clean inputs, predictions show expected uniformity across classes. Right: For triggered inputs, certain classes exhibit unusually high confidence (mean) and multiple classes show near-zero variance, despite the balanced dataset.}
\label{fig:latent_analysis}
\end{figure}

Figure \ref{fig:latent_analysis} shows the mean and variance of model predictions across different classes for both clean and triggered inputs. For clean inputs, we observe that the model's predictions follow a relatively uniform distribution across classes, which is expected given the balanced nature of the dataset. However, when presented with triggered inputs, the model exhibits anomalous behavior: certain classes show unusually high confidence (high mean) in predictions, while multiple classes display near-zero variance in their prediction distributions. This stark contrast in behavior is particularly suspicious given that the dataset has a uniform class distribution.

This observation reveals a critical weakness in latent backdoors. Even during their dormant phase, they process triggered inputs in a distinctly different manner that manifests in the model's output distributions. The presence of highly confident predictions and unnaturally low variances for certain classes, despite the uniform class distribution in the dataset, creates a clear signal that could be exploited for detection. In contrast, as shown in Figure~\ref{fig:deferbad_output}, \texttt{DeferBad} maintains natural output distributions for both clean and triggered inputs during its dormant phase, achieving true concealment of the backdoor.

\section{Analysis of Model Output Distributions}
\label{app:output_analysis}

\begin{figure}[h]
\begin{center}
\begin{subfigure}{\textwidth}
\centering
\includegraphics[width=0.8\textwidth]{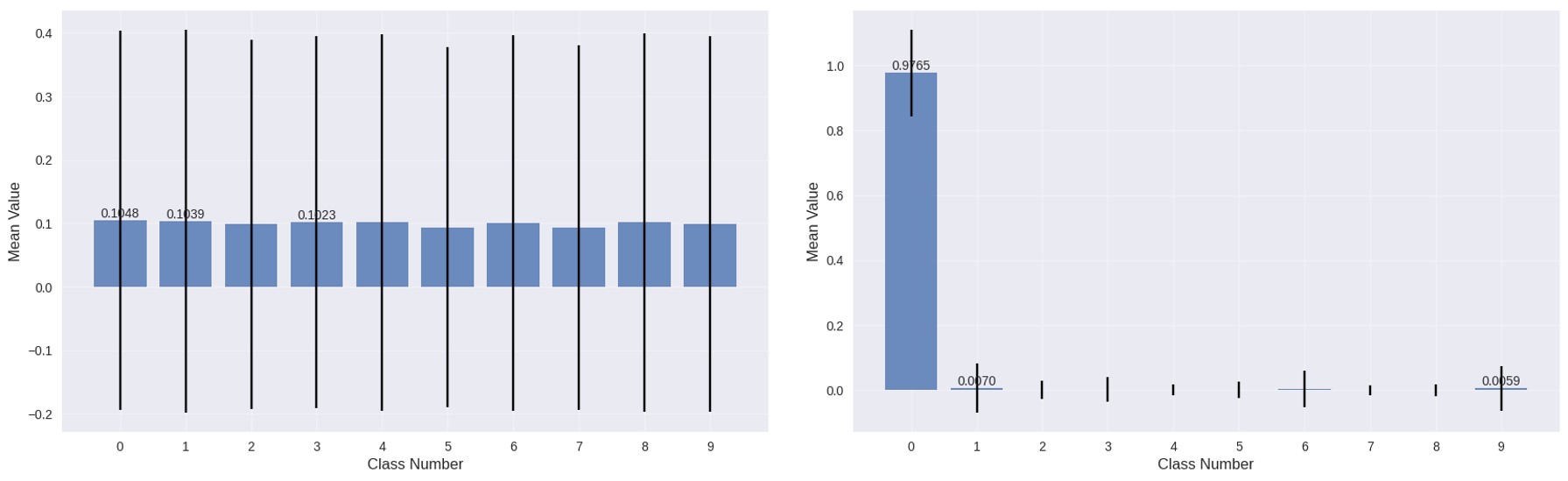}
\caption{Conventional backdoor \citep{badnets} distinctly different patterns between clean and triggered inputs.}
\label{fig:conventional_output}
\end{subfigure}

\begin{subfigure}{\textwidth}
\centering
\includegraphics[width=0.8\textwidth]{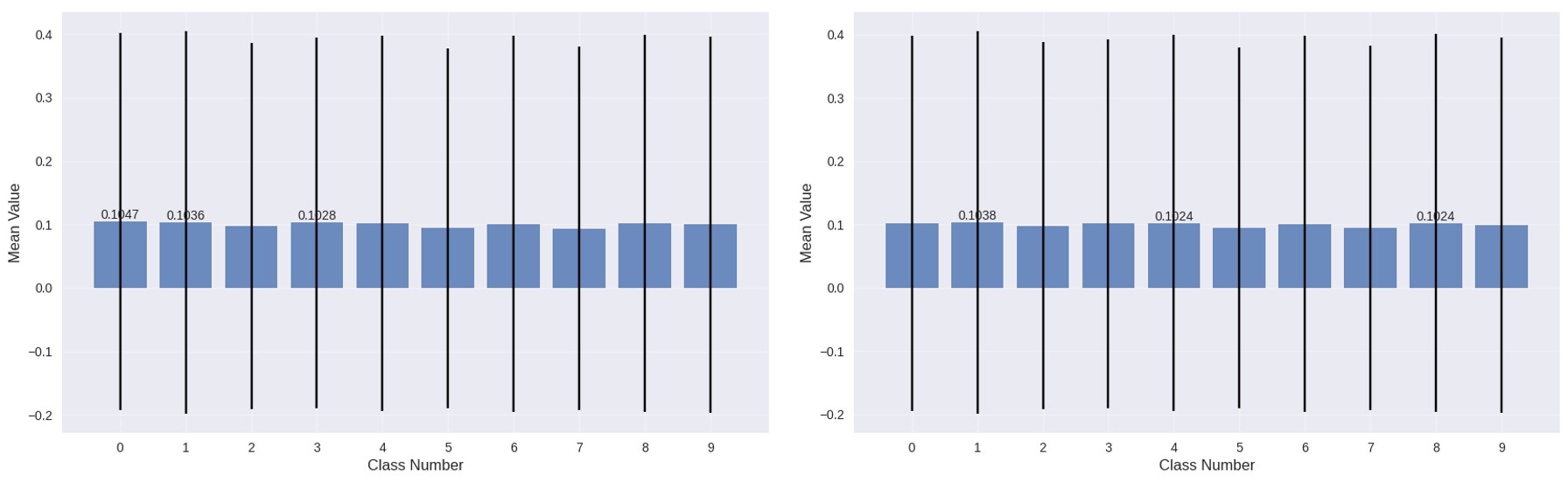}
\caption{DeferBad exhibits nearly identical distributions between clean and triggered inputs.}
\label{fig:deferbad_output}
\end{subfigure}
\caption{Comparison of model output distributions for clean (left) and triggered (right) inputs during the dormant phase. Output distributions are visualized using means and variances across classes.}
\label{fig:output_analysis}
\end{center}
\end{figure}

We analyzed the output distributions of different backdoor approaches during their dormant phase using the CIFAR-10 dataset. Figure \ref{fig:output_analysis} shows the mean and variance of model predictions across different classes for both clean and triggered inputs.

As shown in the figure, conventional backdoors \citep{badnets} produce noticeably different output patterns when presented with triggered inputs, making them potentially detectable through output distribution analysis. In contrast, \texttt{DeferBad} maintains virtually indistinguishable output distributions between clean and triggered inputs, successfully concealing the backdoor's presence.

\end{document}